\documentclass[%
preprint,
 amsmath,amssymb,
 aps,
]{revtex4-2}

\usepackage{graphicx}
\usepackage{dcolumn}
\usepackage{bm}
\usepackage{url}
\usepackage[dvipsnames]{xcolor}

\begin{document}

\preprint{TM-2754-AD-APC-PIP2-TD}

\title{An Upgrade Path for the Fermilab Accelerator Complex}
\thanks{This manuscript has been authored by Fermi Research Alliance, LLC under Contract No. DE-AC02-07CH11359 with the U.S. Department of Energy, Office of Science, Office of High Energy Physics.}%

\author{R.~Ainsworth}
\author{J.~Dey}
\author{J.~Eldred}
\author{R.~Harnik}
\author{J.~Jarvis}
\author{D.E.~Johnson}
\author{I.~Kourbanis}
\author{D.~Neuffer}
\author{E.~Pozdeyev}
\author{M.J.~Syphers}
  \altaffiliation[Also at ]{Department of Physics, Northern Illinois University, DeKalb, IL, 60115, USA.}
  \email{syphers@fnal.gov}
\author{A.~Valishev}
\author{V.P.~Yakovlev}
\author{R.~Zwaska}
\affiliation{%
 Fermi National Accelerator Laboratory, Batavia, IL, 60510, USA
}%

\date{\today}

\begin{abstract}
The completion of the PIP-II project and its superconducting linear accelerator will provide up to 1.2 MW of beam power to the LBNF/DUNE facility for neutrino physics.  It will also be able to produce high-power beams directly from the linac that can be used for lower-energy particle physics experiments as well, such as directing beam toward the Muon Campus at Fermilab for example.  Any further significant upgrade of the beam power to DUNE, however, will be impeded by the limitations of the present Booster synchrotron at the facility.  To increase the power to DUNE by a factor of two would require a new accelerator arrangement to feed the Main Injector that does not include the Booster.  In what follows, a path toward upgrading the Fermilab accelerator complex to bring the beam power for DUNE to 2.4 MW is presented, using a new rapid-cycling synchrotron plus an energy upgrade to the PIP-II linac.  The path includes the ability to instigate a new lower-energy, very high-power beam delivery system for experiments that can address much of the science program presented by the Booster Replacement Science Working Group. It also allows for the future possibility to go beyond 2.4 MW up to roughly 4 MW from the Main Injector.  
\end{abstract}

\maketitle

\newpage

\tableofcontents

\newpage

\section{\label{sec:motivation} Motivation}

\subsection{\label{sec:Science1} Science Opportunities}

\begin{figure}[htp]
\begin{centering}
\includegraphics[height=180pt]{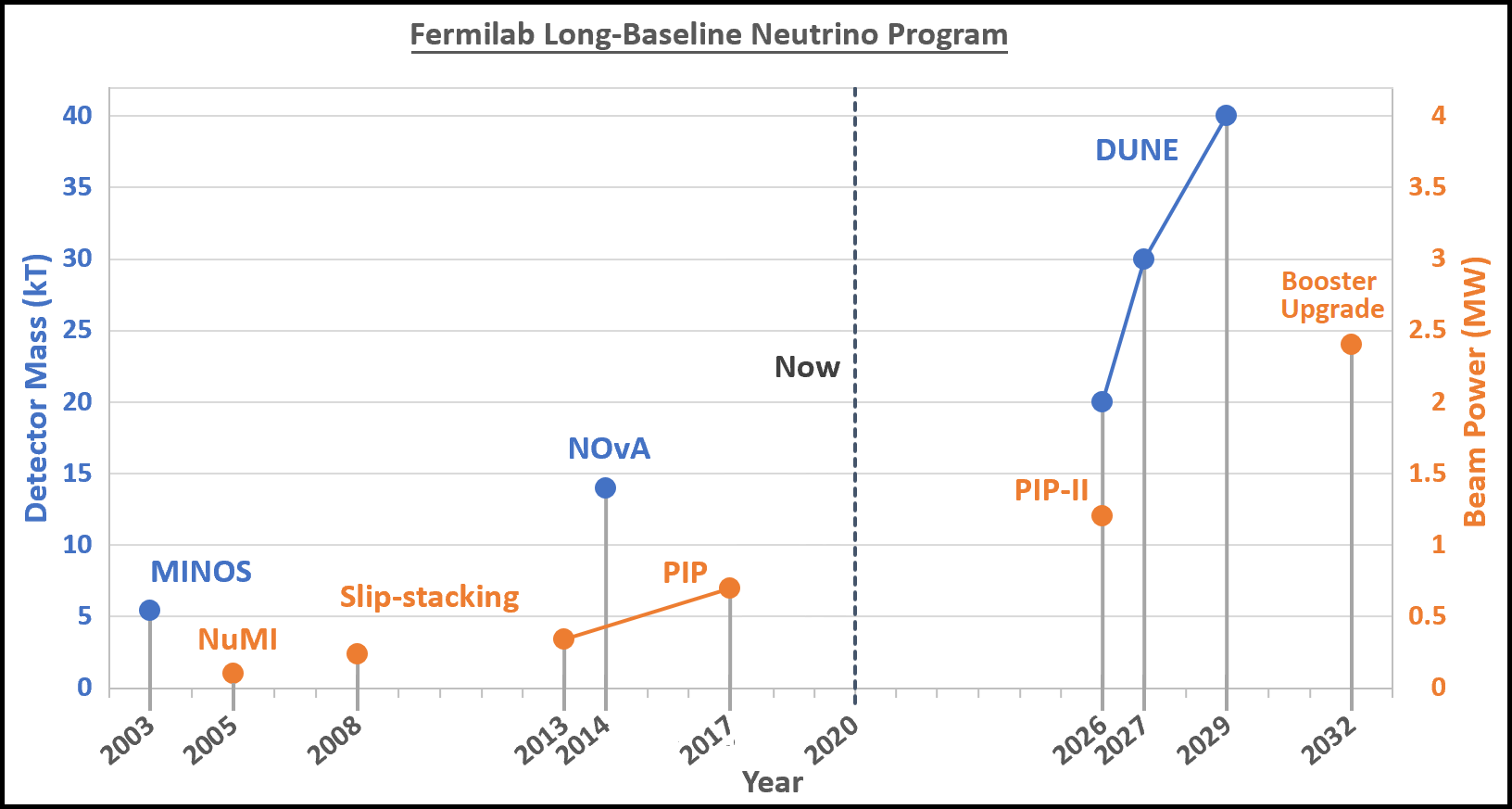}
 \caption{Past and projected milestones in Fermilab long-baseline neutrino program, as measured in detector mass and 120~GeV beam power at the Main Injector.~\cite{Eldred}}
  \label{History}
\end{centering}
\end{figure}

Whenever Fermilab has advanced the scale of its long-baseline neutrino detectors, it has been advantageous to increase proton power to the neutrino source commensurately. Fig.~\ref{History} shows a timeline for detector and accelerator milestones of the Fermilab long-baseline neutrino program.

The next flagship long-baseline neutrino experiment at Fermilab, DUNE/LBNF, constitutes an international multi-decadal physics program for leading-edge neutrino science and proton decay studies~\cite{DUNE}. The Proton Improvement Plan II (PIP-II) is expected to achieve 1.2~MW beam power for the DUNE/LBNF program through the construction of a new 0.8~GeV SRF linac with a series of upgrade and improvements for the Fermilab Booster and Main Injector~\cite{PIP2}.

\begin{figure}[htp]
\begin{centering}
\includegraphics[height=140pt]{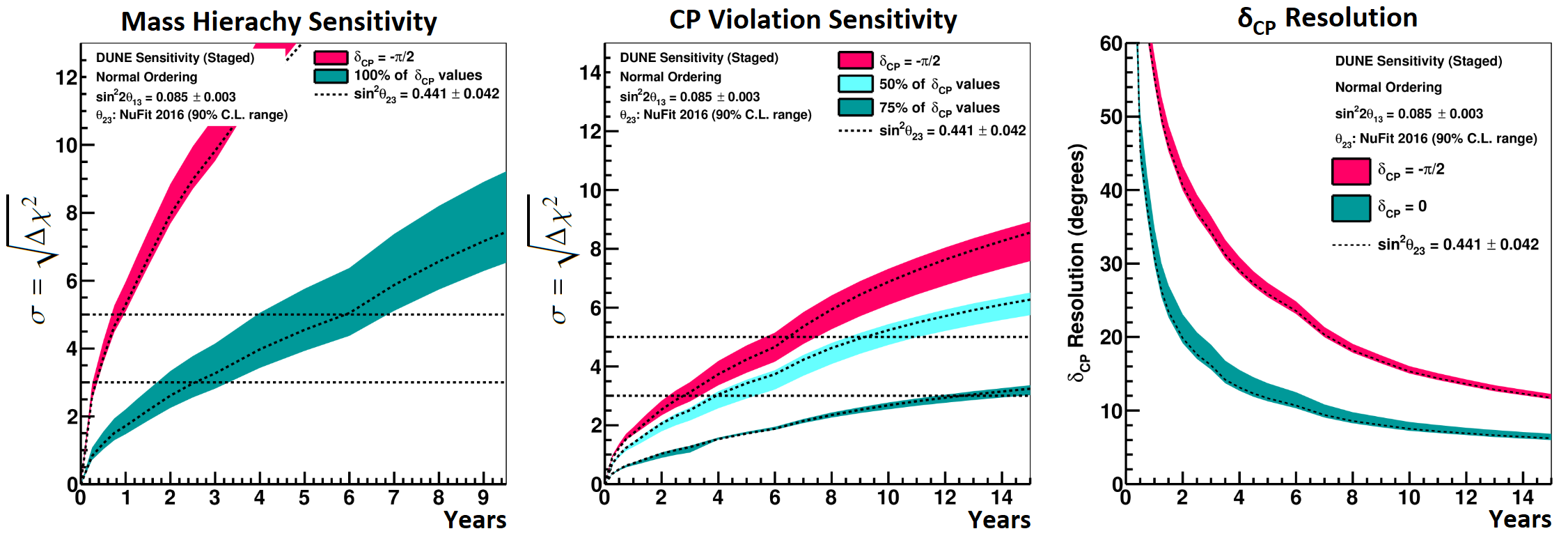}
 \caption{Anticipated DUNE physics results, adapted from \cite{DUNE}, with a 2.4 MW upgrade at the 6-year mark. (left) Sensitivity to determining the neutrino mass hierarchy. (center) Sensitivity to determining a nonzero CP-violating phase. (right) Resolution of the CP-violating phase.} 
  \label{DUNE}
\end{centering}
\end{figure}

The DUNE program has called for a 2.4~MW upgrade of the LBNF beamline in order to achieve its long-baseline physics milestones (as recommended by a 2.4-MW upgrade, DUNE is competitive with and complementary to other long-baseline neutrino experiments proposed on a similar timescale \cite{ICFA,T2K,ESS}.  Figure~\ref{DUNE} shows the anticipated fundamental physics results with the DUNE Technical Design Report deployment scenario.


Protons for the Main Injector are presently supplied by the Fermilab Booster, but the Booster cannot reach the intensity ($13-30 \times10^{12}$) required to achieve 2.4~MW in the Main Injector. After PIP-II, the geometry of the Booster will not accommodate further increases in injection energy. At intensities in excess of the 6.5$\times10^{12}$ required for PIP-II, collective effects may lead to prohibitive losses during transition-crossing. Consequently, the 2.4~MW upgrade must replace the Fermilab Booster with a modern higher-power particle accelerator. Section~II provides an overview of a 2.4~MW upgrade scenario obtained by a combination of a 2~GeV upgrade of the PIP-II linac, a new Rapid-Cycling Synchroton (RCS) to replace the Booster, and RF upgrades to the Main Injector complex to accommodate the higher power. The 2.4~MW upgrade path is also compatible with a subsequent upgrade to 4~MW, through an increase in the Main Injector ramp rate (see Section~\ref{sec:4MW}).

In addition to the long-baseline neutrino program from the Main Injector program, a well-planned accelerator upgrade should give consideration to the physics opportunities available at low and intermediate energies. The 2.4~MW RCS option we discuss in this paper is compatible with MW-class beam power at low ($\sim$2~GeV) and intermediate ($\sim$8~GeV) energies, simultaneous with the Main Injector program (see Section~\ref{subsec:rcs}).

There is a companion white paper~\cite{Harnik} to this document, compiled by the Booster Replacement Science Working group, which surveys the compelling science opportunities available in next-generation physics experiments. We find a wide scope of physics opportunities that may be enabled or brought closer to realization by the PIP-II linac upgrade, the new high-power accelerator, and/or the enhanced power of the Main Injector. The goal of \cite{Harnik} is \emph{not} necessarily to prioritize the potential experiments, perform a cost assessment, or to restrict siting options to Fermilab. Rather the goal is to inform the accelerator design by describing experiments that may be proposed in the years ahead and the special beam requirements needed to pursue them. With this informed accelerator design, we hope that many doors will remain open to pursue exciting physics goals, such as searches for dark sectors, a variety of charged lepton flavor violation searches, and precision measurements. 

In \cite{Harnik}, the science working group has striven to collect possible physics opportunities with concrete options that are feasible in the short term, but also with a long-term vision. The Fermilab Booster, replaced by this upgrade, was designed over 50 years ago and has been utilized well beyond the expectations of its designers. Thinking in the same way with regards to the Booster's replacement, we expect the new machine will serve the HEP community for many decades to come. Versatility and upgradability, therefore, are also core design values.

The structure of \cite{Harnik}, is arranged in many numbered sections each of which present physics opportunities put forth by the community. The topics were discussed in an open virtual workshop on May 19th 2020 and input was collected from the community in the following weeks. Every section contains a brief motivation and physics case, and a description of the experimental setup. Since the goal of that compilation is to inform the accelerator design, each section also contains a subsection that specifies accelerator needs, such as the type of beam, the beam energy and intensity, the needed time structure, etc. Each topic also situates itself in the present context of its global field of study.

The physics topics represent a broad array, pursuing goals in dark sector physics, neutrino physics, charged lepton flavor violation (CLFV), precision tests, as well as R\&D facilities, both for detector development, and to explore new directions for HEP. The list of presented topics is shown in Table~\ref{tab:summary} and labeled in these broad categories. 

\begin{table}[t]
    \centering
    \caption{A summary of the physics opportunities presented in this document, categorized by areas of physics they pursue.}
\includegraphics[width=14.5cm]{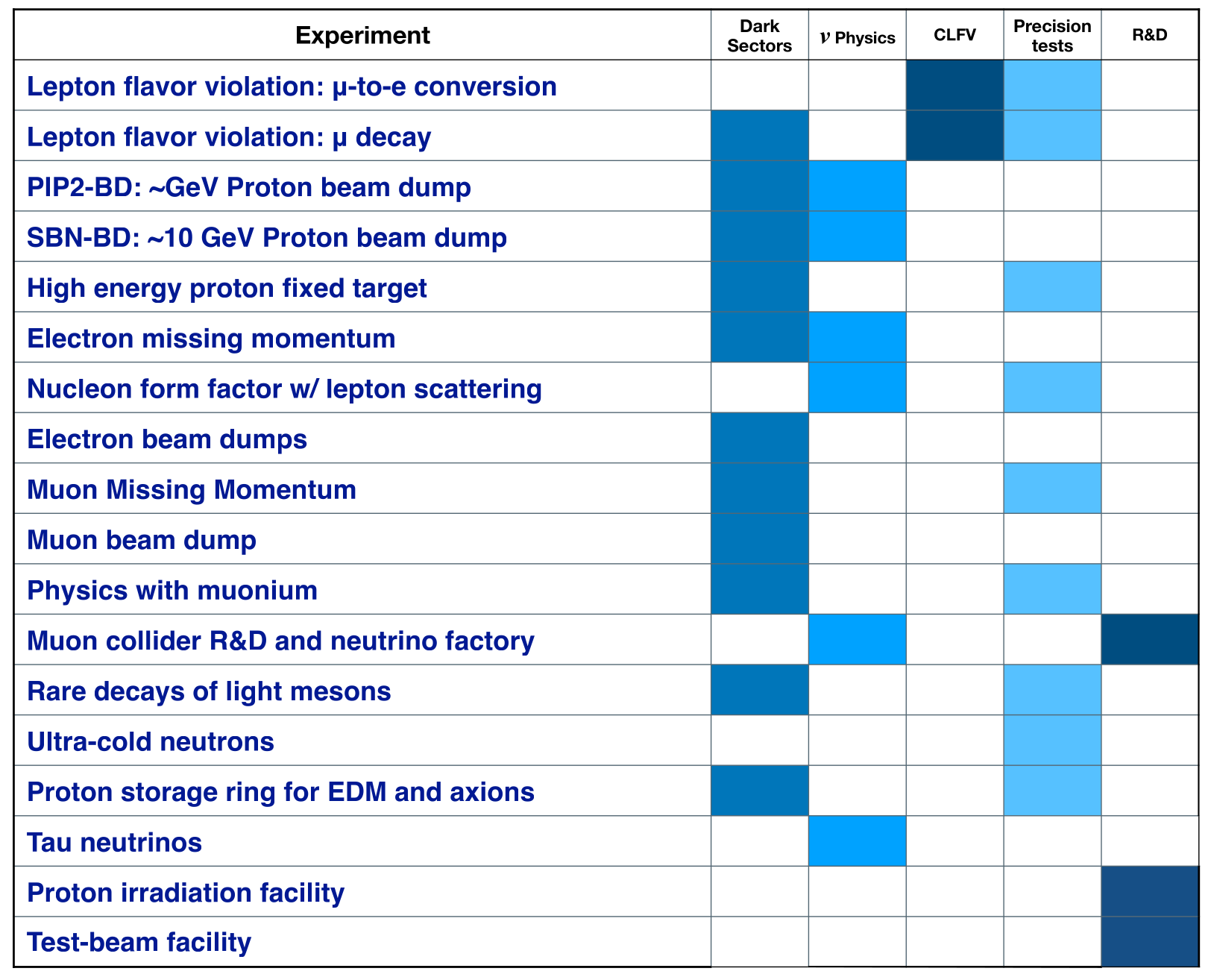}
    \begin{tabular}{c|c}
    \end{tabular}
    \label{tab:summary}
\end{table}

\subsection{\label{sec:Science2} Beams Available for Science}

After the Booster Replacement Science Working group~\cite{Harnik} outlined the broad scope of science opportunities and their individual beam requirements, we outline here how the proposed programs could be powered by the linac and RCS upgrades detailed in the sections below.

The PIP-II linac beam at 0.8~GeV is intended to support continuous beam users, for example the proposed mu2e-II experiment~\cite{mu2e2}. In Section~\ref{energy_upgrade_section}, we describe a 2~GeV extension of the PIP-II linac for injection into the RCS as well as a dedicated experimental program. Proposed experiments using 0.8 to 2~GeV cw proton beams include the mu2e-II CLFV experiment (see \cite{Harnik} Section~2), CLFV muon-decay experiments (Section~3), nucleon form factor by muon scattering (Section~8), the MAGE muonium experiment (Section~12), REDTOP rare-decay program (Section~14), and ultra-cold neutron-antineutron annihilation experiments~(Section~15).

With a new polarized proton injector into the PIP-II linac (Section~\ref{source_upgrades} of this document), the linac could accelerate the polarized beam to an electric dipole moment and axion experiment (\cite{Harnik} Section~16).

Section~\ref{sec:RCSinjection} and \ref{sec:ARlattice} of this document describes how a 2~GeV Accumulator Ring (AR) could facilitate H$^{-}$ injection into the RCS. The same ring would be capable of delivering intense pulses of $\sim$35$\times10^{12}$ protons to a new 2~GeV beamline. The AR pulse~rate would be tightly constrained by a H$^{-}$ foil-stripping injection, but as much as 60 or 120~Hz with H$^{-}$ laser-stripping (i.e. a MW-scale program). The proposed experimental program includes a PRISM-like CLFV experiment (\cite{Harnik} Section~2), as well as a stopped pion and low-energy dark matter program (Section~4). An important R\&D topic will be the performance of pulse-compression methods for experiments, where we might expect to achieve a factor of 4-5 pulse compression while reducing the per-pulse intensity by less than a factor of 2.

Concurrent with the 120~GeV Main Injector program, the RCS described in this document would provide 0.75~MW at 8~GeV (35$\times10^{12}$ protons every 20~Hz cycle). Possible experiments for the RCS beamline include a kaon decay-at-rest and intermediate-energy dark matter search~(\cite{Harnik} Section~5), muon beam dump experiment (Section~11), NuSTORM program and muon collider R\&D (Section~13), proton irradiation facility (Section~18), and/or any successor experiments to the current short-baseline neutrino program~\cite{SBND}. For an experiment that requires continuous beam at intermediate energy, such as the muon beam dump experiment, the muon campus Delivery Ring performing slow-extraction could be used.  Section~\ref{sec:RCSgreater} discusses options for increasing the beam power, pulse rate and extracted energy of the RCS.


The tau neutrino appearance measurement (\cite{Harnik} Section~17) is an aspect of the DUNE/LBNF program available at high energy (120~GeV and optimized horns) and high power ($80-800$ events per MW-yr). Obviously, the entirety of the DUNE/LBNF beam program benefits from the 2.4~MW upgrade, and even further by a subsequent 4~MW upgrade (see Section~\ref{sec:4MW} of this paper).

The Main Injector currently operates a slow-extraction program at 120~GeV, taking 5-7 seconds once every 60 second supercycle, otherwise dedicated to the long-baseline program. Two experimental proposals use or extend these slow-extraction beamlines, namely a high-energy dark-matter search (\cite{Harnik} Section~6), a muon missing momentum experiment (Section~10), and a test beam facility (Section~19). The present Main Injector slow-extraction is loss-limited, and some improvements may be possible with modernized slow-extraction hardware and methods. However the proposed experiments do not require an improvement to slow-extraction efficiency and the 2.4~MW upgrade does not directly impact slow-extraction efficiency.

For completeness, \cite{Harnik} also highlights several experimental programs that use electrons or positrons - an electron missing momentum program (Section~7), nucleon form factor by electron scattering (Section 8), and electron beam dump (Section~9). The 2.4~MW proton facility upgrade described in this document does not directly advance any of these experiments, as the PIP-II p$^{+}$/H$^{-}$ linac cannot easily be re-purposed as an electron/positron accelerator.

\section{\label{sec:upgradepath} Accelerator Upgrade Path}

\subsection{\label{sec:accIntro} Introduction}

Phase II of the Fermilab Proton Improvement Plan (PIP-II) is driven by the need to deliver an average of 1.2~MW of proton beam power at energies of 60-120~GeV per proton on the target to deliver neutrino beams to the DUNE experiment. The new PIP-II superconducting linear accelerator system will replace the present Fermilab Linac as the primary proton source and has the potential to deliver higher intensity proton beams for a future robust science program, both for DUNE and for a lower-energy, high-intensity facility.

To take full advantage of the DUNE program a further doubling of the beam power on target to 2.4~MW is being considered.  Regardless of the specific configuration of the upgraded complex, the most significant bottleneck in such an upgrade is the Fermilab Booster synchrotron which cannot achieve the $13-30\times10^{12}$ protons required for 2.4~MW operation of the Main Injector.

A substantial Booster intensity increase will drive unacceptable space-charge induced losses unless the injection energy can also be increased~\cite{ShiltsevEldred}, but it will not be possible for the Booster to increase injection energy beyond the 0.8~GeV PIP-II upgrade. The length required for an appropriate H$^{-}$ injection straight (at least 1.4~GeV) cannot be accommodated by the Booster magnet layout or even the tunnel geometry. The Booster also faces several unique challenges - including transition-crossing, limited aperture, severe magnet impedances - that would be completely eliminated by a modern machine.

A scenario of achieving 2-MW beam power in the Fermilab Main Injector (MI) by replacing the Booster with a new rapid-cycling synchrotron (RCS) was originally laid out in the 2003 Proton Driver Study II (PD2)~\cite{PDriver}. In 2010, a superseding RCS proposal was described in the Project X Initial Configuration Document 2 (ICD-2)~\cite{ICD2}. An updated concept~\cite{Nagaitsev}, based on the ICD-2 proposal, uses a 2-GeV upgrade of the PIP-II linac with a cost-effective 8-GeV RCS that ramps at 10~Hz and accumulates batches (pulses) in the Recycler. A separate RCS scenario for a 2.4-MW Main Injector, featuring a 1-GeV linac, 15-Hz 11-GeV RCS, and slip-stacking was considered in \cite{Eldred,Eldred2}.  

Additional upgrade paths have been considered that include the possibility of direct injection into the Recycler Ring (RR) or MI from an upgraded 8 GeV linac \cite{ICD2}.  Injection is geometrically constrained to be in the MI/RR-10 region and has been further restricted by the present plan to extract toward DUNE at MI-10.  A separate white paper will revisit this configuration with updates for the current PIP-II and Main Injector geometries and further consideration of any technology R\&D that is required for high-efficiency injection.

What is presented below is one scenario that can achieve 2.4~MW to DUNE while enabling a lower-energy science program that is in line with many of the science opportunities presented in the previous chapter.  While variations of this scenario exist and will no doubt be contemplated in the future, this scenario meets the essential requirements using existing methods and technologies and provides a starting point for more detailed studies and discussions.  Below are the main highlights of the scenario:

\begin{itemize}

\item Central to the scheme will be the construction of a new RCS system to replace the existing Booster.  Although higher energies could be considered, the output energy of the RCS presented here is 8~GeV to minimize the impact on existing Main Injector systems such as the injection infrastructure.  The RCS system will run at a higher repetition rate than the present Booster in order to reduce the fill time into the Main Injector, and it will be optimized in a variety of ways for high-intensity operation.  Additionally, the RCS design will emphasize compatibility with potential technology upgrades, which may enable greater stability and performance at higher intensities.

\item An extension of the PIP-II linac to 2~GeV allows for a higher injection energy into the RCS in order to mitigate space charge effects and also expands the science opportunities with lower energy beams.

\item With the higher repetition rate and higher intensity of the RCS system, the slip-stacking operations performed in the Recycler synchrotron will no longer be required, simplifying the operation and minimizing the overall accelerator time line.  
\item Main Injector power supply improvements will reduce the overall cycle time of the system to under 1~second, improving the average beam power to DUNE.

\end{itemize}

The above approach is meant to optimize the existing infrastructure with a new high intensity accelerator system that can (a) meet the doubling of the beam power for DUNE in a relatively short period of time,  (b) bring to the laboratory a new lower-energy yet very high-intensity scientific program which best utilizes the new system, and (c) provide a path to even higher beam power on target for DUNE.  The scenario as described removes the need of the Recycler as a storage ring.  Hence, once 8 GeV beam is no longer required for the present Muon campus, space for an RF upgrade in the MI tunnel is enabled.  This, coupled with a power supply upgrade, would improve the Main Injector ramp rate, providing even higher beam power to DUNE.  As is typical, exact timing and implementation of all phases of the upgrade will require careful planning and coordination between laboratory management and the experimental program.


\subsection{\label{subsec:protoneconomics} Requirements and Limitations}
The following basic principles were adhered to in arriving at the proposed injector upgrade path:
\begin{itemize}
    \item The accelerator complex development will be based on the capabilities and infrastructure resulting from the implementation of PIP-II.
    \item The upgrade shall be balanced with respect to the practicality of implementation on one side, and the potential for further development on the other.
    \item The upgrade scenario will be based on well-established technologies.
    \item The upgrade to the complex shall enable reliable and safe operations. The beam losses shall be kept at a level preventing excessive component and enclosure activation to allow for safe and efficient accelerator maintenance. 
\end{itemize}







\subsection{Synchrotron and Linac Options}

Previous studies of Fermilab upgrade possibilities have considered Linac-based and RCS-based alternatives, and discussions of each implementation continue.  Below is a list of potential advantages of an RCS-based option:
\begin{itemize}
    \item This may provide the most straightforward replacement of the existing RCS injector, namely the Booster.  Injection, accumulation and other operational procedures in the MI could remain substantially unchanged.
    \item The pulsed delivery of beam is desirable for many experiments, particularly neutrino beams.  An RCS readily re-packages linac beam as pulsed beam, although the pulse structure is constrained by the fixed ring circumference.
    \item A new RCS could have an injection section designed and optimized for H- injection.  Losses at injection would occur at lower energy (2 GeV) rather than the 8 GeV of a full-energy Linac injection. 
\end{itemize}


And following is a list of potential advantages of a Linac-based option:
\begin{itemize}
    \item Many experiments benefit from a CW beam, particularly lepton flavor conservation experiments. This is most readily obtained from a CW linac. 
    \item While foil-stripping injection may be more difficult at 8 GeV into the existing Recycler Ring or Main Injector, laser-assisted injection may be easier because lower-frequency lasers can be used.  The required laser power at these lower frequencies will be much more easily obtained.
\end{itemize}

\begin{figure}[h]
\begin{centering}
\includegraphics[height=90pt]{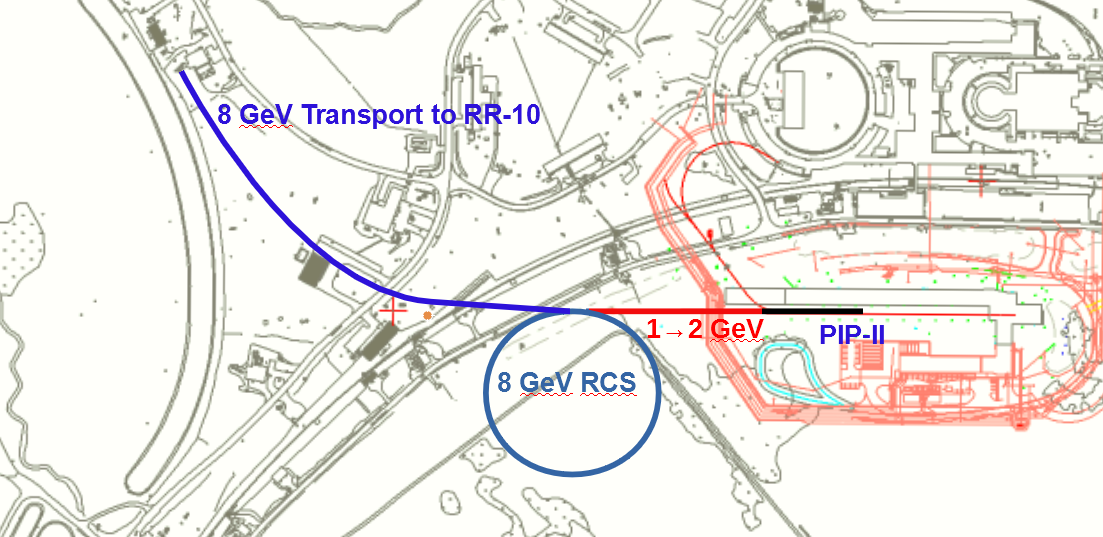}
\includegraphics[height=90pt]{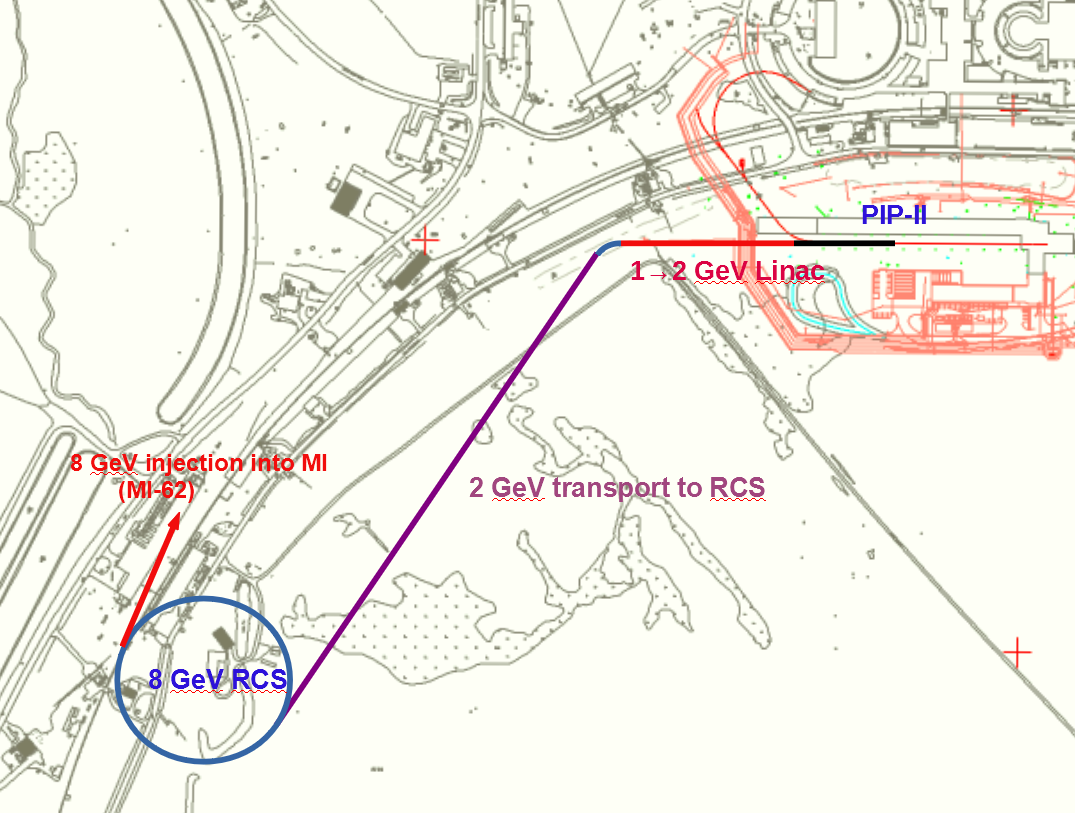}
\includegraphics[height=90pt]{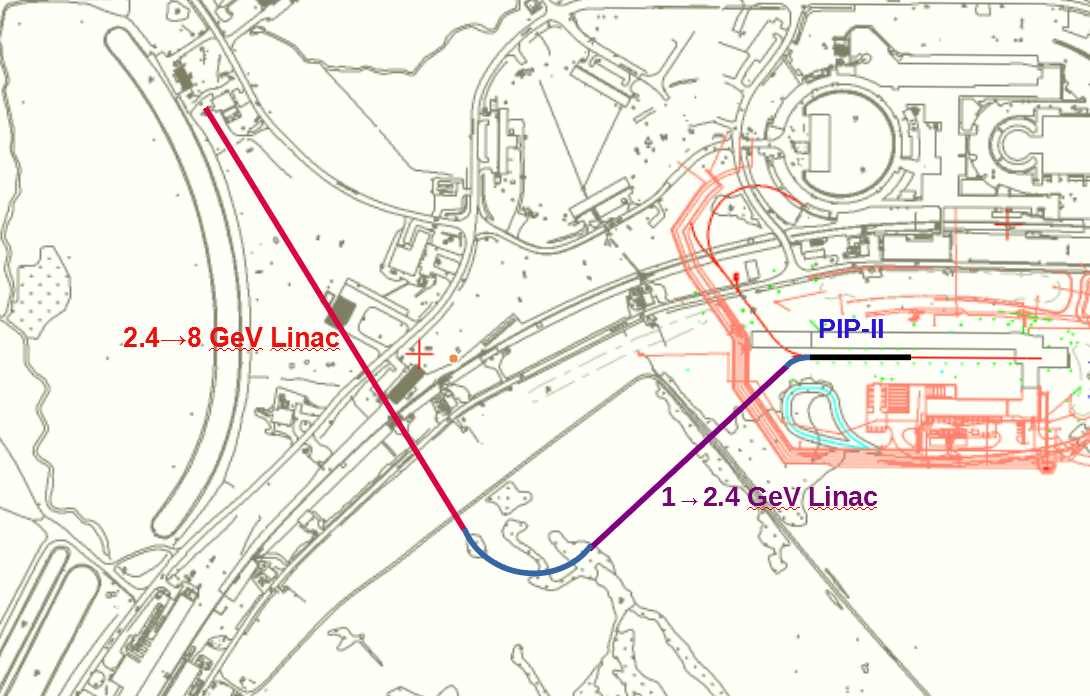}
 \caption{Some Booster Replacement Upgrade options.  Left:   Linac upgrade plus RCS, tying into existing 8 GeV beam line system.  Center:  Linac upgrade with RCS, injecting into MI62 straight section in Main Injector.  Right:  Linac-only upgrade.} 
  \label{fig:upgradeOpts}
\end{centering}
\end{figure}

The linac scenario produces intrinsically higher duty factors at 8~GeV for experiments, while the RCS scenario produces intrinsically lower duty factors. A small array of possible configurations of linac and RCS upgrade options are presented in Fig.~\ref{fig:upgradeOpts}.  

\subsection{An Upgrade Scenario}

This accelerator design exercise has been an attempt to deliver a possible upgrade path that in principle can satisfy the near-term goal, on a reasonable time scale, of doubling the beam power to DUNE beyond the PIP-II expected performance level  while enabling directions for even higher power beam to DUNE as well as a possible lower-energy science program at Fermilab.   Several options for replacing the bottleneck system -- the Booster synchrotron --  can be contemplated, including the extension of the PIP-II linac energy to 8 GeV or a direct replacement of the Booster itself with a more modern system, for example, or a combination of both.  The option of replacing the Booster with a new rapid-cycling synchrotron (RCS) has received extensive investigation and the scenario presented meets the goals of the study in a straightforward way.  This is not to say that new concepts and other scenarios cannot meet the same goals if more effort were invested into those directions.  But with the scenario at hand, a direction for the laboratory can be envisioned wherein the high-energy neutrino program can be upgraded well beyond the expected PIP-II era levels, while enabling a new, high-power lower-energy world-class science program at the laboratory utilizing much of the existing infrastructure of the present injector system.

The scenario that will be discussed in the upcoming sections of this report involves two major upgrades, namely an extension to the PIP-II linac increasing its output energy, and the replacement of the Booster with a new RCS system to deliver high intensity proton beams directly to the Main Injector.  The exact output energy of the linac is partially determined through input from the scientific community, in an attempt to maximize the amount of science that can be performed with this lower-energy high power beam. The output beam energy of the RCS for this study has been kept at 8~GeV for optimization of performance within the existing injector complex, but could also be fine-tuned to a different value (10, or 12~GeV, say) if a compelling science case were made.

\begin{figure}[h]
\begin{centering}
\includegraphics[height=280pt]{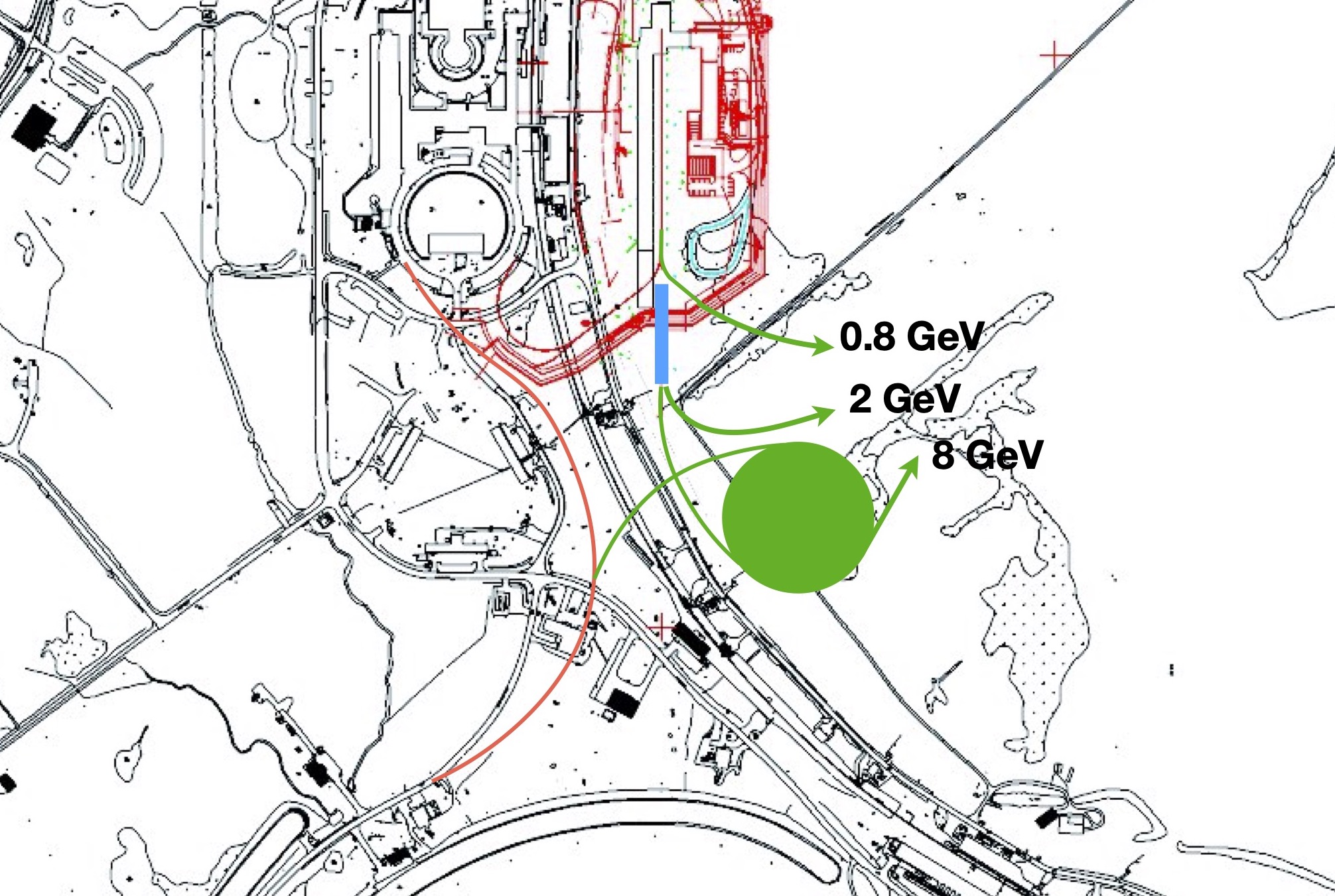}
 \caption{Possible Booster replacement upgrade layout on Fermilab campus. Blue indicates an extension to the PIP-II linac; green indicates a new rapid cycling synchrotron facility.} 
  \label{fig:overview}
\end{centering}
\end{figure}

Figure~\ref{fig:overview} shows a drawing of the Fermilab site plan with overlays of possible locations for an RCS and an extension of the PIP-II linac.  All layouts of beam lines and accelerators are conceptual only. The ``arrows'' indicate the possible delivery of beam toward a new experimental area from the RCS or directly from the linac system.    

The general parameters of the Booster replacement scenario are presented in Table~\ref{BooUpParams}, with further details presented in the following sections.  The potential beam powers listed in the table are for dedicated operation at the specific energy, assuming a CW linac extension.


\begin{table}[h]
\centering
\caption{General Parameters}
\begin{tabular}{|| l || r | l ||}
\hline
{\bf Parameter} & {\bf Value} & {\bf Unit} \\
\hline\hline
Linac final energy (kinetic) & 2 & GeV \\
Linac ave. current & 2 & mA \\
Linac pulse length & 3 & ms \\
Linac extension length & 120 & m \\
Potential CW beam power at 0.8 GeV & 1.6 & MW \\
Potential CW beam power at  2  GeV &  4 & MW \\
\hline
RCS final energy (kinetic) & 8 & GeV \\
RCS intensity & 35 & $10^{12}$ \\
RCS batches to MI &  5 &  \\
RCS circumference & 570 & m \\
RCS bend field (inj/ext) & 0.31/1.0 & T \\
RCS RF frequency (ext) & 52.8 & MHz \\
RCS RF voltage & 1.2 & MV \\
Potential pulsed beam power$^*$ at 8 GeV & 0.75 & MW \\
\hline
MI intensity &  175 & $10^{12}$  \\
MI cycle time & 1.4  & s \\
MI RF peak voltage & 4.8 & MV \\
MI beam power (120 GeV)  & 2.4  & MW \\
Potential beam power with ramp upgrade & 4 & MW \\
\hline
\end{tabular}

$^*$ Concurrent with 2.4 MW MI operation.
\label{BooUpParams}
\end{table}

\subsection{LBNF Beamline Upgrades \& Targetry Capabilities}

A 2015 technical description of the LBNF beamline can be found in the DUNE/LBNF CDR~Vol.~3~\cite{LBNFcdr3} and Annex~3a~\cite{Annex3A}. The LBNF facility is designed to be compatible with 2.4~MW beam power operation. In 2017, the target and  geometry was optimized for physics reach at 1.2~MW under mechanical constraints by genetic algorithm~\cite{Fields}, and the LBNF beamline was been updated~\cite{Annex3Aopt}. Beam operation of the facility beyond 1.2~MW will require upgraded designs of the target, and some portion of the horns and beam windows. A possible 4~MW upgrade of the DUNE/LBNF program would likely require second target hall.


The design of the LBNF target itself was informed by the helium-cooled T2K carbon target as well as NuMI operational experience. In an Aug 2019 conceptual design review, a cantilevered target design was selected for 1.2~MW operation out of three options, with preference towards minimizing target exchange downtime and technical risk~\cite{Densham}. Future efforts would include designs of a 2.4~MW target, horns, windows, and other devices, as well as the integration plans for retrofitting the facility.

Target lifetime and indicators of target integrity are also an active area of R\&D work. The Radiation Damage In Accelerator Target Environments (RaDIATE) collaboration~\cite{Senor} was developed to study a broad suite of accelerator target, dump, window, and collimator materials under extensive radiation damage and thermal shock. As part of the RaDIATE collaboration, accelerator relevant materials are exposed to high radiation dose at the Brookhaven Linac Isotope Producer (BLIP) facility, after which they can be subject to examination and testing procedures. CERN's High-Radiation to Materials (HiRadMat) facility enables the post-irradiation materials to subsequently be exposed to thermal shock testing from a high-energy accelerator beam. Presently, Fermilab has a 400~MeV irradiation test area, ITA, and Section~18 of \cite{Harnik} contains a proposal consistent with an 8~GeV proton irradiation facility that would deliver $10^18$ protons within a few hours; other options for irradiation at Fermilab include target stations fed by PIP-II for radiation damage, and the AP-0 target station for thermal shock. Fermilab is also constructing a Target Systems Integration Building which will enable the assembly of new devices for LBNF, and include a High-Power Targetry Laboratory for the investigation of the properties of irradiated materials and assemblies, either already in use or proposed.

\subsection{Accelerator Staging}

With the development of a new accelerator system to replace the present Booster it is imperative that the construction be performed in a way to minimize the impact on on-going operations.  With the upgrade as described, a staged approach can be envisioned, though details remain to be worked out.  For instance, one could envision the following scenario:

\begin{itemize}
   \item Commission PIP-II for 1.2 MW operation to the DUNE program.
   \item Enable an 800 MeV science program from PIP-II.
   \item While running the above programs, 
\begin{itemize}
    \item build linac extension to 2 GeV,
    \item commission the 2 GeV linac extension using spare linac pulses in the operational time line, and
    \item commission 2 GeV science program; meanwhile,
    \item construct the RCS and associated beam lines, and
    \item commission the RCS using low-intensity beam from the 2 GeV linac system.
\end{itemize}
   \item In a somewhat longer shutdown, the Main Injector RF system upgrade would be performed, and the connection of the RCS system with the Main Injector would be completed. 
   \item Commission the RCS/MI system to 2.4 MW to the DUNE program.
   \item At some point, while running high-power low-energy program, during a long shutdown an upgrade to the MI power supply system can be performed to achieve $>$2.4 MW to DUNE.  Prior to the shutdown, service building upgrades and other necessary infrastructure enhancements could be performed.
\end{itemize}



Depending upon scheduling, one can imagine a scenario in which the conventional RCS is constructed prior to the completion of the full energy upgrade of the superconducting PIP-II linac to 2~GeV. The question is, at what injection energy can the linac deliver beam to the RCS and still create 1.2 MW power to DUNE?  To answer this, 
a study was undertaken to look at the performance of the desired RCS layout but with variable injection energy.   
In this model, the RCS intensity is scaled down by keeping the space-charge tune spread and normalized emittance constant. At energies less than 1.2~GeV the geometric beam emittance becomes constrained by beampipe aperture (about 50\% larger geometric emittance than the 2~GeV beam), and so at these energies intensity and normalized emittance are scaled down by keeping the space-charge tune spread and geometric emittance constant.

\begin{figure}[htp]
\begin{centering}
\includegraphics[height=180pt]{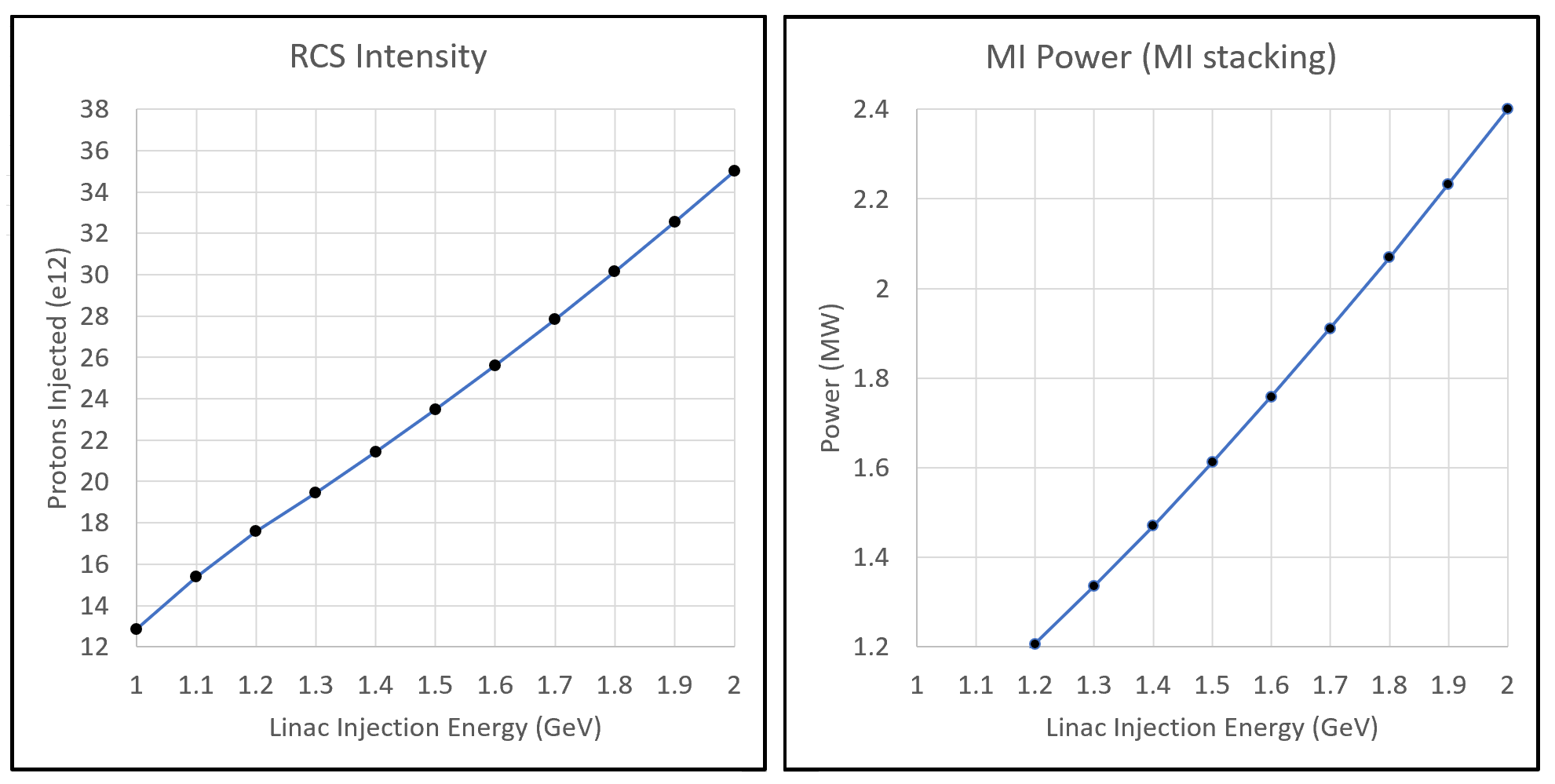}
 \caption{Left, estimated RCS Intensity as a function of linac injection energy. Right, corresponding Main Injector power at 120~GeV, assuming stacking in the Main Injector.}
  \label{Staging}
\end{centering}
\end{figure}

Fig.~\ref{Staging} shows the model of the expected RCS intensity and corresponding Main Injector power. At 1.2~GeV, the Main Injector reaches the 1.2~MW PIP-II benchmark except without relying on the Recycler Ring or slip-stacking. Another natural break point is 1.6~GeV, where the Main Injector achieves the maximum beam power limit (1.8~MW) without upgrading Main Injector RF power. For our RCS design, the intensity required for a 2.4~MW DUNE program is not easily achievable without the full injection energy of 2~GeV. An RCS design to achieve the required intensity at less than 2~GeV is certainly possible, but a full trade-off study between linac cost, RCS cost, and low-energy experimental programs is beyond the scope of this document.

\newpage


\section{\label{subsec:linac} PIP-II Linac}
\subsection{\label{PIP2_linac_section} PIP-II Linac Design and Parameters}

The Proton Improvement Plan-II (PIP-II) encompasses a set of upgrades and improvements to the Fermilab accelerator complex aimed at supporting a world-leading High Energy Physics program over the next several decades. The P5 report \cite{2014p5} recommendation and the DOE-approved Mission Need Statement \cite{pip2mns} define the primary goals for PIP-II: 

\begin{itemize} 
\item Deliver beam with a power of 1.2 MW to the LBNF/DUNE target, upgradable to multi-MW
\item Deliver a platform capable of high-duty-factor/high-beam-power operations and providing flexible bunch patterns to multiple experiments simultaneously
\item Deliver a platform to support future upgrades of the accelerator complex
\item Ensure sustained high reliability of the Fermilab accelerator complex
\item The above capabilities should be provided in a cost-effective manner.
\end{itemize}

\begin{figure}[htp]
\begin{centering}
\includegraphics[height=200pt]{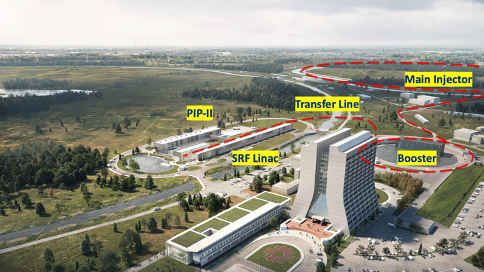}
 \caption{PIP-II Linac location on Fermilab campus.} 
  \label{pip_2_campus}
\end{centering}
\end{figure}

\begin{figure}[htp]
\begin{centering}
\includegraphics[height=200pt]{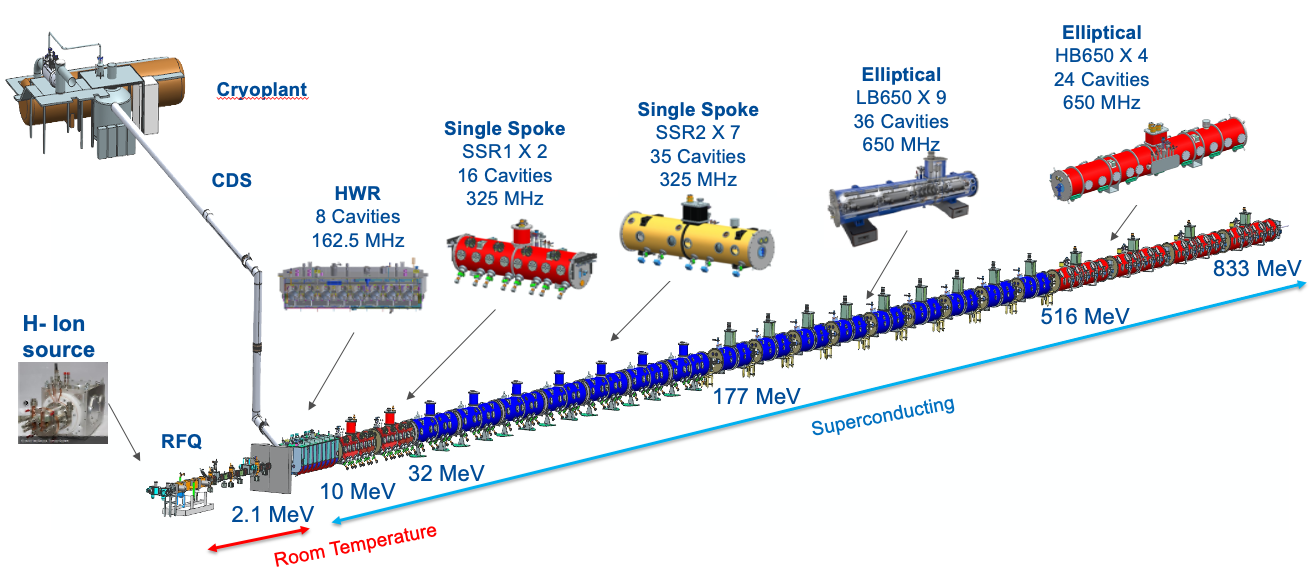}
 \caption{PIP-II Linac with the cryoplant and the cryo distribution line (CDC). The linac consists of the room-temperature front end, one HWR cryomodule, two types of Single-Spoke resonators (SSR1 and SSR2), and two types of elliptical resonators (LB650 and HB650).} 
  \label{pip2_linac}
\end{centering}
\end{figure}

As part of the project, PIP-II delivers a superconducting Linac to fuel the next generation of intensity frontier experiments. The linac will accelerate H- ions to 800 MeV for injection into the Booster. The project includes upgrades to the existing Booster, Main Injector, 
and Recycler rings that will enable them to operate at an increased repetition rate (Booster at 20 Hz and Main Injector at 0.83 Hz) and deliver a 1.2 MW proton beam on the Long Baseline Neutrino Facility (LBNF) target. Figure \ref{pip_2_campus} show the PIP-II Linac with the Beam transfer Line to the Fermilab Booster on the campus. Figure \ref{pip2_linac} shows the layout of the major linac components. Table \ref{PIP2Parameters} lists the major linac parameters. 

\begin{table}[htp]
\centering
\caption{PIP-II Parameters for LBNF and multi-user modes. Corresponding beam parameters in the multi-user mode must be adjusted according to user's share.}
\begin{tabular}{|| l || c || c || l ||}
\hline
Parameter & LBNF Mode & Multi-User Mode & comment \\
\hline
Beam Energy (MeV) & 800  & 800 & upgradable, see \ref{energy_upgrade_section} \\
Ave. current (mA) & 2    & 2   & limited by amplifiers, see \ref{intensity_upgrade_section} \\
Pulse length (ms) & 0.5  & Programmable, CW  &   \\
Bunch rep. rate (MHz) & $\sim70$ & Up to 162.5 &  \\
Min. Bunch spacing (ns)  & 6.2   & 6.2  &   \\
Bunch length (ps) & 4    & 4  & \\
H- per bunch      & $1.9\times10^8$  &  Up to $4\times 10^8$ & \\
H- per pulse      & $6.5\times10^{12}$ & Programmable & \\
Pulse rep. rate (Hz)     & 20   & Adjustable & \\
Beam power (kW)   & 17   & Up to 1600 & \\       
\hline
\end{tabular}
\label{PIP2Parameters}
\end{table}

\subsection{\label{Main PIP-II Linac Systems}Main PIP-II Linac Systems}

\begin{itemize}
    \item \textbf{Ion sources and LEBT}. The baseline design of the PIP-II linac includes two identical multi-cusp, filament-driven, H- sources with their own LEBT branches. A 3-way switching magnet allows switching between sources within minutes. Switching between the two ion sources allows maintaining one of the sources while the other can provide beam for operations, increasing beam availability. The ion sources are designed to operate in the DC regime, producing up to approximately 15 mA of H- beam current.
    \item \textbf{RFQ} The PIP-II RFQ is a 4-vane brazed structure operating at 162.5 MHz. The RFQ is designed to accelerate the beam to 2.1 MeV in the CW regime. The beam dynamics of the RFQ was optimized for acceleration of beam with intensity up to 15~mA. Further increase of the beam intensity leads to beam quality degradation and losses. The RFQ sets the minimum temporal separation between bunches in the linac. 
    \item \textbf{MEBT Chopper} The MEBT chopper is designed to selectively remove bunches without affecting neighboring bunches. The chopper can create arbitrarily programmed bunch patterns. The chopper consists of two kickers, a beam absorber, and pulse forming electronics. The beam intensity is reduced by approximately 60\% in the MEBT by the fast MEBT chopper and collimators to ensure lossless injection of the 162.5 MHz PIP-II bunch pattern into the 45 MHz RF of the Booster.
    \item \textbf{SRF Linac} The SRF linac is designed to accelerate H- beams and consists of five different types of cavities as show in Fig. \ref{pip2_linac}. The number of cavities and their design are optimized to match the velocity profile of the accelerated H- beam. The scope of the PIP-II project includes four HB650 cryomodules, sufficient to accelerate the beam to 833 MeV. The linac tunnel allocates space for additional two HB650 CMs that can accelerate the beam to 1050 MeV. The design of the RF cavities is optimized for CW operations with a beam intensity of several milliamperes. The cavities do not include HOM dampers, as they are not required at this intensity.
    \item \textbf{RF Amplifiers} The maximum beam current accelerated by cavities, assuming the acceleration voltage profile is fixed, is determined by the available RF power. PIP-II Linac RF amplifiers were specified to accelerate beam with a peak current of 2 mA. The PIP-II amplifiers are solid state, CW amplifiers. The power ratings of the PIP-II baseline design amplifiers are 7 kW, 7 kW, 20 kW, 40 kW, and 70 kW for HWR, SSR1, SSR2, LB650, and HB650 cavities respectively. The power specification includes overhead to compensate for transmission losses and to provide control margin. 
    \item \textbf{Beam Transfer Line (BTL)}. The PIP-II project includes a beam transfer line (BTL) to deliver the beam from the linac to the Fermilab Booster. The BTL is approximately 300 m long. It is designed to transport 1 GeV beam and consists of two achromatic arcs, one small and one large, and two straight sections. The BTL also includes a dump rated for 25 kW of beam power. 
\end{itemize}

This choice of systems parameters determines some important PIP-II Linac features:

\begin{itemize}
    \item The MEBT chopper can provide arbitrary, programmable bunch patterns, including gaps and reduced frequency. This functionality is critical for PIP-II operations. Because the frequency of the linac RF is not a harmonic of the Booster RF, a significant portion of the linac bunches will be lost at injection into the Booster. The chopper will selectively remove bunches that would be injected too close to the separatrix or miss the bucket, eliminating losses \cite{pip2pdr}. Based on simulations of the injection process into the Booster, the chopper will have to remove up to 60\% of the bunches. Similar functionality will be required in the case of the RCS. In addition, the chopper will be used to produce the flexible bunch patterns required for users and machine operations, e.g. reduce the bunch frequency or instantaneously turn off and turn on the beam while switching the beam between users to reduce losses.
    \item The highest bunch frequency is 162.5 MHz, determined by the RFQ. The bunch frequency can be reduced using the MEBT chopper. 
    \item The number of H- per bunch can reach to $4\times 10^8$ without suffering significant degradation of the beam quality. The nominal bunch intensity in the LBNF mode is $1.9\times 10^8$ H- per bunch. 
    \item The average beam current (over 1~$\mu$s) is limited to 2 mA ($1.25\times 10^{16}$ H-/sec) by the RF power available from the RF amplifiers. Any combination of bunch frequency and charge is possible if the average current in the pulse does not exceed 2 mA, the bunch frequency does not exceed 162.5 MHz, and the maximum number of particles per bunch does not exceed $4\times 10^8$. Options for increasing the beam current are discussed later in Section \ref{intensity_upgrade_section}.
    \item High quality, low halo beam.
    \item Capable of accelerating protons with the addition of a proton ion source (presently not in scope). See more in Section \ref{source_upgrades}.
    \item The linac is not suitable for acceleration of electrons or ions heavier than protons.
\end{itemize}

\subsection{PIP-II Design Is Compatible With Science Driven Upgrades}

\begin{figure}[htp]
\begin{centering}
\includegraphics[height=180pt]{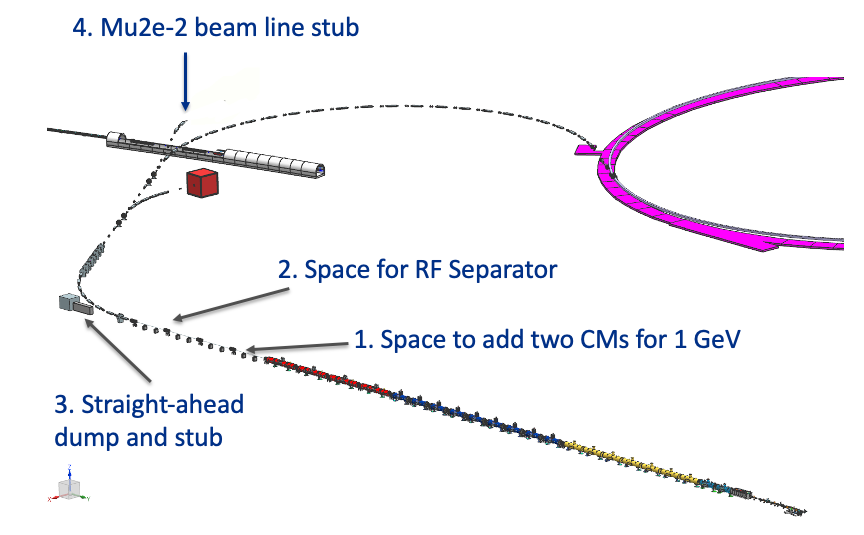}
\caption{Provisions included in the baseline design of the PIP-II linac and BTL to facilitate future upgrades and addition of users.} 
\label{PIP2_Upgrades}
\end{centering}
\end{figure}

The PIP-II Mission Need Statement (MNS) requires PIP-II to deliver 1.2 MW of the beam power onto the LBNF target. The MNS also emphasizes the need to implement PIP-II in a manner that will allow a subsequent doubling of beam power delivered from the Main Injector and maintain compatibility with subsequent upgrades in support of a broader spectrum of particle physics research opportunities. The design of the PIP-II linac includes provisions that facilitate future upgrades and addition of users. Figure \ref{PIP2_Upgrades} shows the layout of the baseline design with design features included to facilitate future upgrades:
\begin{enumerate}
    \item The linac tunnel includes the space and provisions for required infrastructure (e.g. wave guide penetrations) for two additional HB650 cryomodules. The addition of two cryomodules in the linac tunnel will increase the beam energy above 1 GeV.
    \item Space at the downstream end of the linac to add either an RF separator or a beam switchyard with a fast switching magnet. 
    \item A stub at the end of the linac tunnel for straight-ahead extension of the linac to increase the beam energy beyond 1 GeV and provide beam to other experiments and accelerators.
    \item A stub in the beam transfer line to the Booster to provide beam to the Muon Campus and other users. 
\end{enumerate}

\subsection{\label{energy_upgrade_section} Energy Upgrade}

{\it 1 GeV upgrade.}  
The PIP-II linac tunnel includes space for two more HB650 cryomodules, as shown in Figure \ref{PIP2_Upgrades}. The linac beam energy will be increased above 1 GeV by adding two HB650 cyomodules. The conventional facilities include provisions for the systems required to operate the cryomodules, e.g. penetrations for RF wave guides and magnet current lead cables. The cryoplant capacity is sufficient to operate the additional cryomodules. 

{\it 2 GeV upgrade.}
The energy of the PIP-II beam can be further increased by extending the linac tunnel and adding more cryomodules. To facilitate this construction, a stub will exist at the end of the tunnel as shown in Fig. \ref{PIP2_Upgrades}.  Also, the power of the beam dump at the end of the linac is limited to reduce activation of the area at the end of the linac. 

\begin{figure}[htp]
\begin{centering}
\includegraphics[height=180pt]{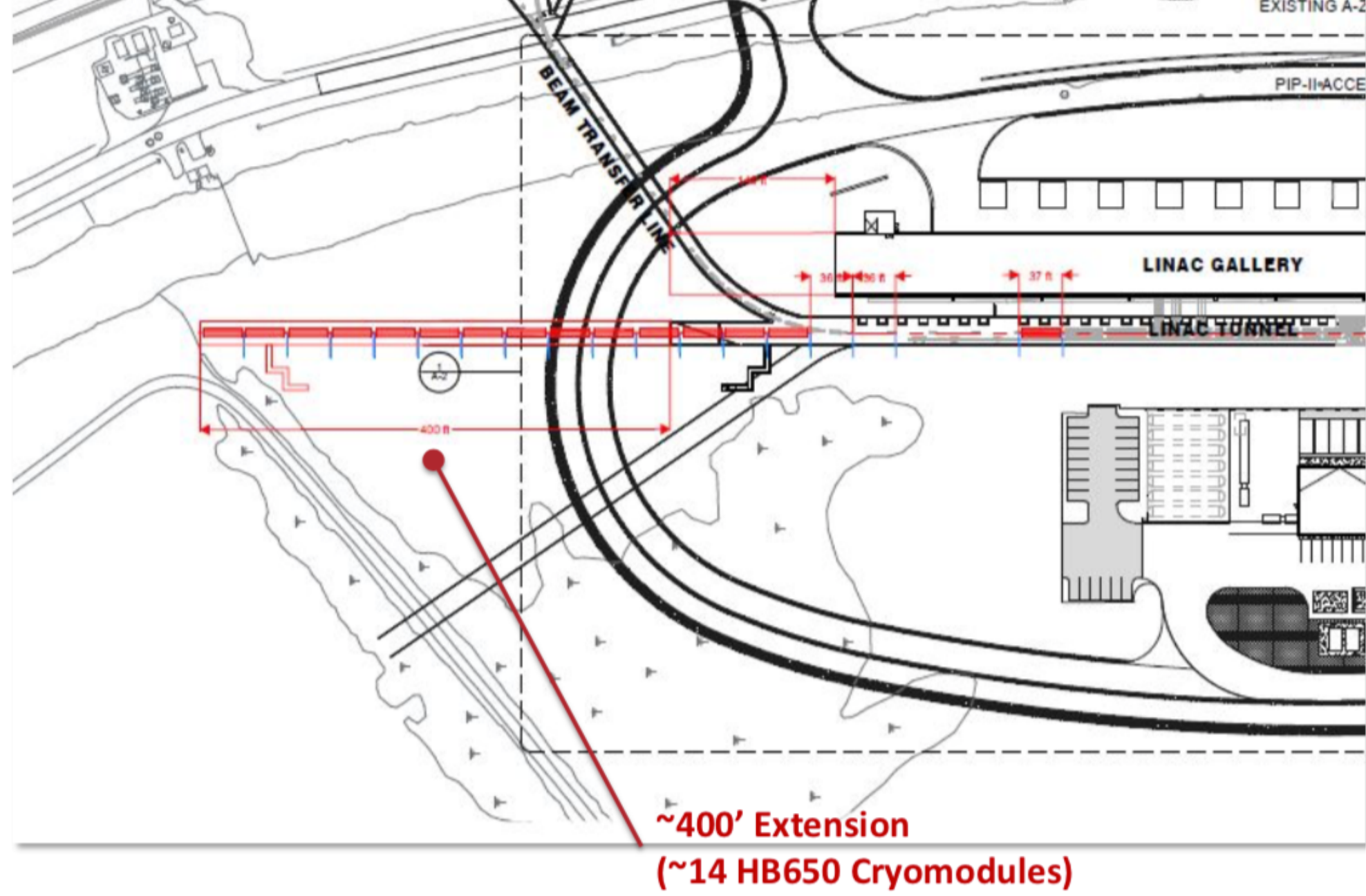}
 \caption{PIP-II Linac extension includes up to 14 HB650 CMs sufficient to reach energy above 2 GeV.} 
  \label{LinacExtension}
\end{centering}
\end{figure}

The linac tunnel can be extended by 400 feet without requiring DOE to conduct a new Environmental Assessment. Figure \ref{LinacExtension} shows the linac extension with the added cryomodules. This extension is sufficient to accommodate 14 more HB650 cryomodules based on the geometrical dimensions of the cryomodules, quadrupoles, and drifts between cryomdules. These 14 cryomodules with the two additional cryomodules in the existing linac tunnel (the 1 GeV option) are expected to increase the beam energy roughly to 2.5 GeV. This beam can be injected into the RCS as described in this paper and/or provided to other users in a concurrent multi-user mode. It is important to note that a smaller number of cryomodules can be operated if a lower energy is required for the RCS and users. Also, the energy upgrade can be staged by gradually adding cryomodules and increasing the beam energy. If CW or high duty-factor modes of operation are not required, pulsed-RF design and mode operation can be considered for the upgrade, enabling significant cost reductions for the cryoplant, RF systems, infrastructure, etc. 

Previous studies conducted to compare the efficiency of HB650 and  1.3 GHz (TESLA) cryomodules for the linac extension concluded that 650 MHz HB650 cavities and cryomodules were more efficient below 2 GeV, and possibly as high as 3 GeV. By the time the PIP-II energy upgrade will be ready to be implemented, the design and performance of the HB650 cryomodule and its cavities will be validated and significant experience with their manufacturing, testing, and operation will be obtained. 

\subsection{Beam Switching options and Multi-User Operations}

The PIP-II linac (with no energy upgrade) is capable of delivering a CW beam with a power of 1.6 MW. The LBNF/DUNE experiment in the PIP-II era (1.2 MW on LBNF target) requires only approximately 1.1\% of the total beam intensity. Even with the doubled beam power on target, the LBNF beam will require only 2.2\% of the linac CW intensity. The rest of the beam can be delivered to multiple users, enabling concurrent operations. 

There are two main types of devices that can be used to distribute beam to multiple users. 

\begin{figure}[htp]
\begin{centering}
\includegraphics[height=180pt]{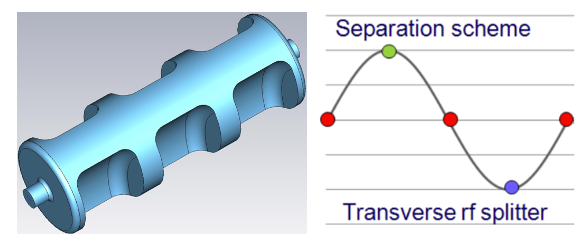}
 \caption{RF Separator Concept. An RF cavity, operating at a harmonic number of the bunch repetition frequency, is designed to provide a transverse kick. The amplitude and the direction of the deflection angle depend on the phase of bunch arrival in the cavity.} 
  \label{RFSeparator}
\end{centering}
\end{figure}

\begin{itemize}
    \item \textbf{RF Separators} are RF cavities that operate at a harmonic number of the bunch repetition frequency and separate the beam in two or more beamlets. Instead of accelerating the beam, RF separators are designed to provide a transverse kick using either electric or magnetic RF field. Figure \ref{RFSeparator} shows the principle of operation of an RF separator. The kick provided by the separator depends on the bunch arrival phase. This separator will split the CW beam with a frequency of 162.5 MHz into three beamlets: two 40.125 MHz CW beams pointing up and down and one 80.5 MHz CW beam going through without deflection. The RF separator requires a drift after the cavity to increase separation between bunch trajectories sufficiently to insert a septum magnet. Using the fast chopper in the MEBT, some bunches can be removed, enabling operations with only two beamlets. In all of these cases, the sum current of all the beamlets cannot exceed the maximum average linac beam current of 2 mA. Note that RF separators are used successfully at many accelerator facilities, e.g. CEBAF at the Thomas Jefferson National Accelerator Facility. 
    \item \textbf{Fast Switching magnets} can deflect the beam to required experiments and switch between beam destinations in 10-20 microseconds. Such a magnet can be programmed to switch the beam periodically between multiple users in a quasi-concurrent manner, delivering periodic bursts of beam with the full pulse intensity. The fast MEBT chopper will turn off the beam in the linac during magnet switching to avoid beam losses.
\end{itemize}

\begin{figure}[htp]
\begin{centering}
\includegraphics[height=180pt]{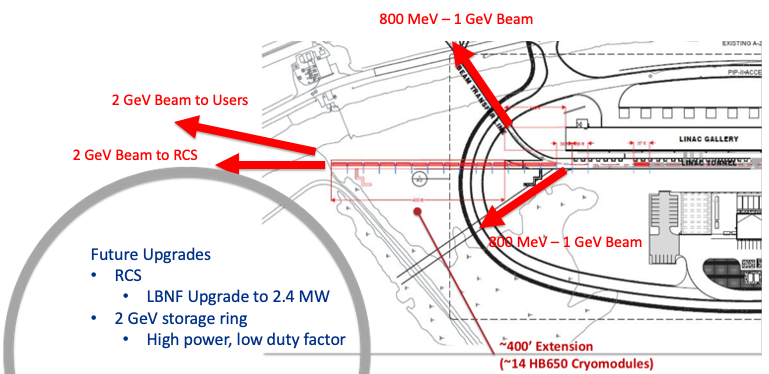}
\caption{PIP-II Linac extension to 2 GeV into the RCS. with extraction lines at 2 GeV and 1 GeV. } \label{MultiUser}
\end{centering}
\end{figure}

RF separators and fast switching magnets can be combined in any required combination making PIP-II capable  of  providing  flexible  bunch  patterns and  high  duty  factor/higher beam power operations to multiple experiments simultaneously. Figure \ref{MultiUser} shows possible beam distribution options for concurrent operation of multiple users. In these options, either a 3-way fast switching magnet or an RF separator at the end of the PIP-II linac would direct the 0.8 -- 1 GeV beam to Muon campus/FNAL Booster, other users, and to the extension of the linac for further acceleration to a higher energy. The beam can be further divided downstream using additional RF separators and fast switching magnets. Note that an RF separator might require a superconducting cavity with the relevant infrastructure to provide a required deflection angle at a high energy.

\subsection{\label{intensity_upgrade_section} Intensity Upgrade}

To achieve 1.2 MW of beam power on the LBNF target, PIP-II needs to accelerate 550 $\mu$s-long pulses with an average peak current of 2 mA. This peak beam intensity is an order of magnitude lower than that in some other pulsed, high-power accelerators such as SNS or ESS.  This choice provides several advantages. First, it reduces intensity-dependent effects and alleviates their impact on the beam quality, simplifying beam chopping and allowing for precision painting during injection into the Booster. Second, it allows using widely available, easy-to-operate, CW-capable solid state amplifiers. On the other hand, a lower beam intensity requires longer injection into RCS. Injecting long pulses in RCS can be problematic due to injection foil overheating and the requirement to keep the RCS field flat during injection. Increasing the beam current in the linac can mitigate these issues. It is conceivable that other future experiments can benefit from increased beam current as well. 

In considering options for boosting the linac intensity, we divide the intensity increase into two ranges loosely based on the impact on accelerator systems and beam parameters: a moderate increase by roughly a factor of 2 and a substantial increase by a factor of 5 to 10. 

\textbf{Moderate increase of linac beam current.} Results of numerical studies and engineering estimates show that increasing the beam current to 4-5 mA, that is, by a factor of 2 to 2.5 relative to the PIP-II baseline design, is feasible without significant design changes and requires only increasing the output power of RF amplifiers roughly proportionally to the beam current. The increased RF power output can affect requirements for the facility electrical power, utilities, and space. Therefore, more compact and efficient amplifier designs shall be evaluated as an alternative to solid state amplifiers. 

\textbf{Substantial increase of linac beam current.} An increase of the linac beam current by a factor of 5 to 10 (from 2 mA to 10 mA to 20 mA) will have a significant impact on accelerator performance. This impact needs to be fully evaluated. Results of these studies can require optimization of machine parameters, operational regimes, and, ultimately, the machine design to mitigate effects caused by the increased intensity. Below is a list of systems and issues that require close attention. 

\begin{itemize}
\item \textbf{Ion source.} 
Note that the beam current extracted from the ion source has to be a factor of 2.5 times higher than the beam current in the linac (see section \ref{PIP2_linac_section}). Thus, the ion source will need to produce 25 to 50 mA of beam current. Presently, no existing H- high-brightness source can produce this much current in the DC regime. There are pulsed state-of-the-art H- sources than can operate with this pulse intensity. For example, the SNS H- source can reliably generate 40-50 mA, 1 ms-long beam pulses at 60 Hz with a duty factor of 6\%. 

\item \textbf{Accelerator front end}.
The front end plays a critical role in the accelerator. Besides generating the beam and providing initial acceleration, it is responsible for forming the beam, including the beam quality and the bunch pattern. The significantly increased space charge will adversely affect performance of the front end and, as a result, the whole accelerator.

The PIP-II RFQ was designed for a maximum current of ~15 mA, corresponding to a current of 6 mA in the linac. Increasing the beam intensity beyond these values will cause significant emittance growth, reduced transmission in the RFQ, and losses in the linac. Although these issues can be partially mitigated by more aggressive collimation in the MEBT, a new RFQ with the design optimized for a higher current will likely be required. 

The beam transport in the front end needs to be reevaluated in the presence of strong space charge. The strength and, possibly, the arrangement of focusing elements will need to be optimized to compensate the defocusing effect of the space charge and prevent degradation of the beam quality and halo formation. 

The effectiveness of the MEBT chopping scheme with the significantly increased beam intensity needs to be verified. The distorted optics and deteriorated beam quality can reduce the efficiency of the fast MEBT chopper, increasing beam losses in the linac and at injection in the RCS to an unacceptable level. 

In mitigating high intensity effects in the front end, it is important to optimize the front end as an integrated system that defines the beam in the machine. The design of several critical systems and a choice of their operational parameters need to be re-examined.

\begin{itemize}
\item Ion source
\item RFQ design, including its frequency and energy
\item Beam transport optics in LEBT and MEBT 
\item The beam chopping scheme
\end{itemize}

These changes can affect the low energy part of the SRF linac and the choice of parameters for the first few cryomodules. A room-temperature structure, such as the DTL, can be considered as an alternative to the first few cryomodules, especially, if the linac can be operated in a pulsed regime.

\item \textbf{SRF cavities and couplers.} 
PIP-II cavities were designed for a low beam current and high gradient. The position of the fundamental coupler was optimized for low coupling and might not be suitable for high current operations. Thus, it is necessary to demonstrate that PIP-II SRF cavities can be optimally coupled for the significantly increased beam current. In addition, it is necessary to ensure that PIP-II couplers are compatible with the required power, both average and peak. Finally, PIP-II cavities do not include High Order Mode (HOM) couplers or other provisions to damp the field of HOMs. It will be necessary to demonstrate that the field excited by the beam in undamped HOMs cannot adversely affect cavity performance and beam parameters (longitudinal and transverse beam instabilities). 

\item \textbf{RF Amplifiers.} The output RF power of the amplifiers needs to be increased proportionally with the beam current, reaching the 350 kW - 700 kW range for HB650 cavities. This output power will require amplifiers of a different design operating in a pulsed regime. 

\item \textbf{Intrabeam stripping.} Intrabeam stripping of H- ions causes beam losses that potentially can limit beam intensity. The rate of intrabeam stripping scales quadratically with the bunch intensity. At approximately 6-7 mA, the local beam loss due to intrabeam stripping will reach 1 W/m, assuming other beam parameters are unchanged. The beam peak intensity can be further increased if the beam duty factor is decreased to keep average losses under control. Thus, intrabeam stripping can prevent high-current CW operations and limit beam availability to users.
\end{itemize}

Thus, technical solutions and operational parameters adapted for the design of the PIP-II Linac will need to be reexamined. Very likely, the linac will require significant modifications to be able to operate with 10-20 mA beams. The changes can affect the design of the front end, including the lower part of the SRF linac, the design of cavities and couplers, the choice of the bunch frequency, and the operational mode (pulsed instead of CW). 

As can be seen from the preceding discussion, substantial increases to beam intensity in the PIP-II linac system will require much study and investment.  For our upgrade scenario envisioned below, only modest increases in beam intensity already achievable by the PIP-II system, on the scale of factors of 2-3 or less, are contemplated.

\subsection{\label{source_upgrades} Source Upgrades}

\begin{figure}[htp]
\begin{centering}
\includegraphics[height=120pt]{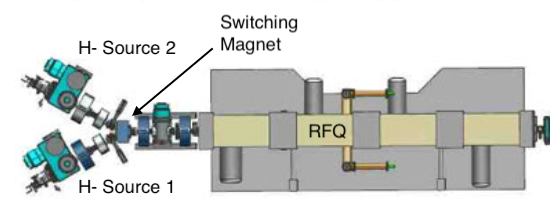}
\caption{Two PIP-II ion sources, LEBT, and RFQ} 
\label{PIP2_sources_baseline}
\end{centering}
\end{figure}

Although the PIP-II linac was designed to accelerate H- ions, it is also capable of accelerating protons.  Most of the accelerator components, such as the RFQ, quadrupole magnets, solenoids, and cavities, are capable of accelerating protons without modifications or changing their polarity and/or phase. The components that need further evaluation and possibly adjustments to their design and operational procedures are LEBT and MEBT choppers. Also, instrumentation that relies on electron stripping, such as laser-wire profile monitors, will not be able to operate with protons. 

The LEBT 3-way magnet can be used to inject protons into the RFQ simultaneously with the H-, thus, enabling simultaneous acceleration of proton and H- beams in the linac. As an option, protons can be accelerated between H- LBNF pulses. A simple dipole magnet at the end of the linac will separate the beams. Simultaneous acceleration of H- and proton beams will require careful evaluation of the orbit correction procedure because dipole correctors in the linac will steer H- and proton beams in opposite directions. 

\begin{figure}[htp]
\begin{centering}
\includegraphics[height=180pt]{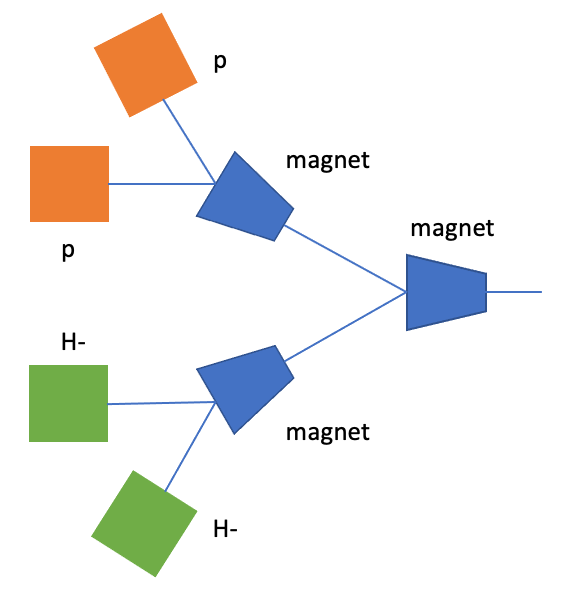}
\caption{Two PIP-II ion sources, LEBT, and RFQ} 
\label{hminus_proton_sourcces}
\end{centering}
\end{figure}

Using a proton ion source instead of one of the H- sources as shown in Fig. \ref{PIP2_sources_baseline} will impact beam availability to LBNF and other experiments due to required source maintenance. To avoid this proton sources can be added by means of additional merging/switching dipole magnets as shown in Fig. \ref{hminus_proton_sourcces}. The impact of the longer LEBT transport combined with the beam space charge on the beam quality needs to be carefully evaluated. 

Conceptually, it is feasible to operate the accelerator with two different H- sources by switching between sources sequentially using a fast electrostatic deflector. The switching time can be sub-microsecond. Conceivably, this option can be employed if two H- sources have significantly different properties, e.g. one of the sources is polarized. This option, although it is feasible and simple compared to the complexity of rest of the project, requires significant redesign of the LEBT.

\section{\label{subsec:rcs} Rapid Cycling Synchrotron}
\subsection{\label{sec:RCSbroad} Broad Parameters} 
The RCS parameters described in Table~\ref{Param} enable 2.4~MW Main Injector operation and a nearly MW-scale 8~GeV beamline program. The Main Injector program is upgradeable to 4~MW and the 8-GeV beamline program to $\sim$1.2~MW (but not both concurrently).

\begin{table}[htp]
\centering
\caption{RCS General Parameters.}
\begin{tabular}{|| l || c || l ||}
\hline
Parameter & Value & Unit \\
\hline
RCS Intensity & 35 & $10^{12}$ \\
RCS Injection Energy & 2 & GeV \\
RCS Extraction Energy & 8 & GeV \\
RCS Normalized Emittance (95\%) & 24 & mm~mrad \\
RCS Tune-Shift & 0.20 & \\
RCS Vertical Aperture & 4.4-6.2 & cm \\
\hline
RCS Circumference & $\leq$ 570 & m \\
Number of RCS batches & 5 &  \\
RCS Ramp-Rate & 20 & Hz \\
Available RCS Power (concurrent with 60 GeV MI) & 0.65 & MW \\
Available RCS Power (concurrent with 120 GeV MI) & 0.75 & MW \\
\hline
MI Intensity & 175 &  $10^{12}$ \\
MI Power (60 Gev MI) & 1.85 & MW \\
MI Power (80 Gev MI) & 2.00 & MW \\
MI Power (120 Gev MI) & 2.40 & MW \\
\hline
\end{tabular}
\label{Param}
\end{table}

For an 8~GeV extracted RCS energy, much of the existing Booster-MI transfer line infrastructure can be re-purposed (see Section~\ref{subsec:mi}).

At 0.8~GeV the RCS would have an extreme space-charge tune-shift of -0.62, but the tune-shift can be suppressed down to -0.2 by upgrading the PIP-II linac energy to 2~GeV (described in Section~\ref{subsec:linac}). The RCS lattice design should also be superperiodic to enhance dynamic aperture. A superperiodic RCS lattice design would facilitate an RCS design which can (optionally) be made compatible with integrable optics or electron-lens technology~\cite{IOTA1,IOTA2,EldredI} which would act to mitigate collective instabilities and halo formation.

To minimize uncontrolled losses and emittance growth, the RCS would feature phase-space painted injection, modest space-charge, aggressive collimation, and transition-free acceleration. The RCS can use a metalized ceramic beampipe to prevent eddy-current heating. If the Fermilab proton complex maintains the 53~MHz RF frequency structure currently in use, the harmonic number of the RCS is 101.

\subsection{\label{sec:RCSinjection} H$^{-}$ Injection} 


A nearly 3~ms injection time is required to fill the RCS with the 2~mA PIP-II linac beam, which presents two challenges. The first challenge is that the bend field changes by 1-2\% over the course of the injection time due to the resonant-circuit ramping magnets. The second challenge is foil-stripping injection for 3~ms at high-energy, which requires a long injection straight, with large beta functions and high-power beam collimators to control foil temperature and beam scattering. The large number of injection turns greatly increases foil heating.

Retrofitting the PIP-II linac as a 5-10~mA pulsed linac would alleviate both of these challenges with the multi-ms fill time. Section~\ref{subsec:linac} describe the required upgrades.


Alternatively, a 2~GeV accumulator ring (AR) would also alleviate both challenges associated with the multi-ms fill time, but also expand the range of science programs that can be accommodated. The AR could use permanent or DC-powered magnets, share the same tunnel with the RCS, and/or have wider apertures and longer straight sections. The 2~mA beam could also be foil-injected using several 120~Hz painting cycles and then transferred to the RCS for immediate acceleration.

\begin{figure}[htp]
\begin{centering}
\includegraphics[height=180pt]{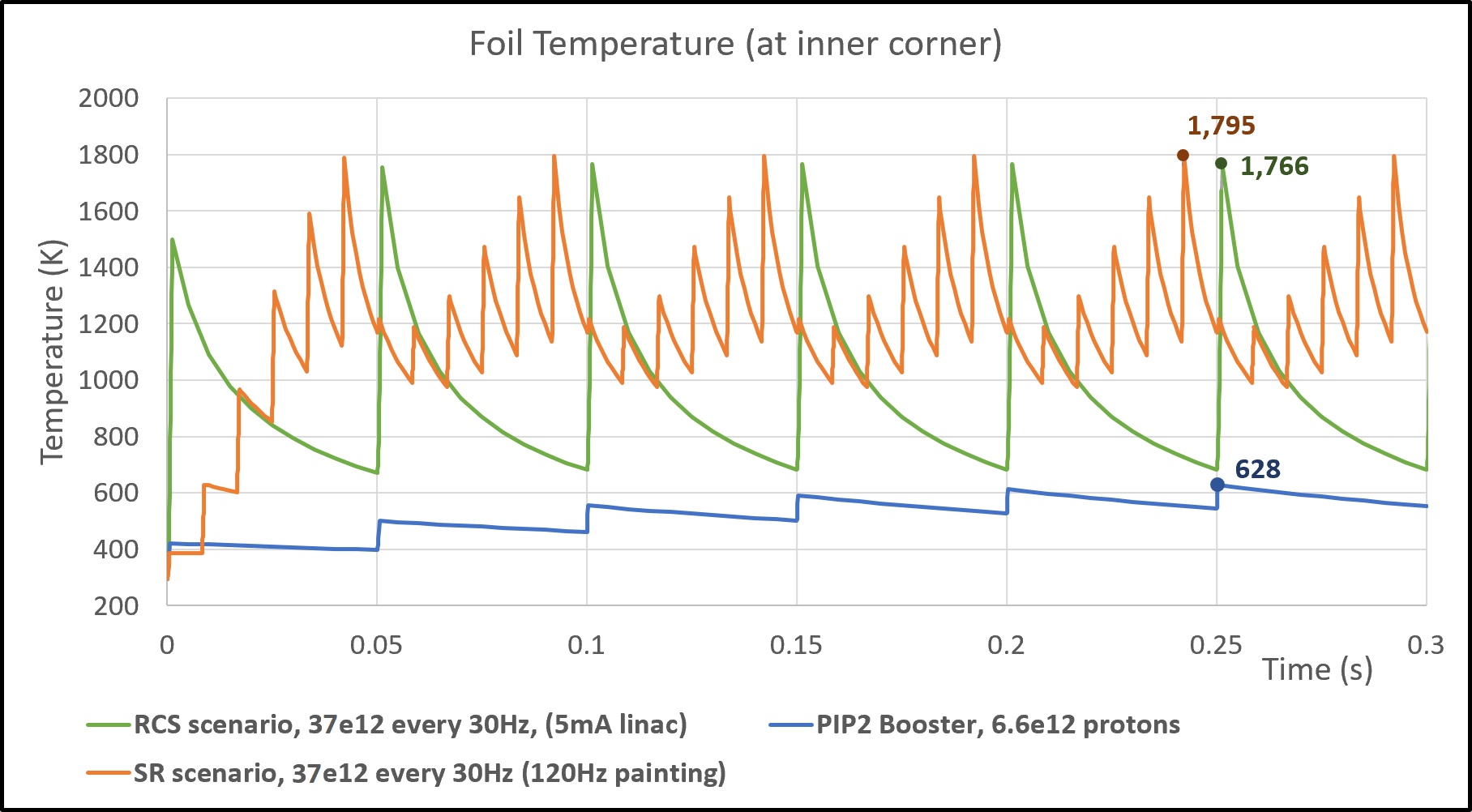}
 \caption{In green, the injection foil temperature in the RCS with beam from 5~mA linac. The injection time is 1.2~ms. In orange, the injection foil temperature in the 2~GeV accumulator ring with 120~Hz painting cycling cycles. (Six 0.5~ms injections every 20~Hz cycle.) In blue, the injection foil temperature in the PIP-II Booster.} 
  \label{Foil}
\end{centering}
\end{figure}

For H$^{-}$ stripping injection by foil, the foil will be heated by the circulating proton beam over the injection time, and the peak foil temperature can be calculated using the method outlined in~\cite{Eldred}. Figure~\ref{Foil} shows the foil heating calculation for two RCS scenarios in comparison to the PIP-II Booster foil heating \cite{PIP2}. In the first scenario, the PIP-II linac is retrofitted for 5~mA pulsed beam injected directly into the RCS. In the second scenario, 2~mA PIP-II linac beam is injected into an accumulator ring, in six 120~Hz painting pulses to fill the ring every 20~Hz. Both scenarios assume $35\times 10^{12}$ protons, 2~GeV injection energy, and preliminary parameters for individual foil thickness and beam optics   .

\begin{figure}[htp]
\begin{centering}
\includegraphics[height=200pt, width=280pt]{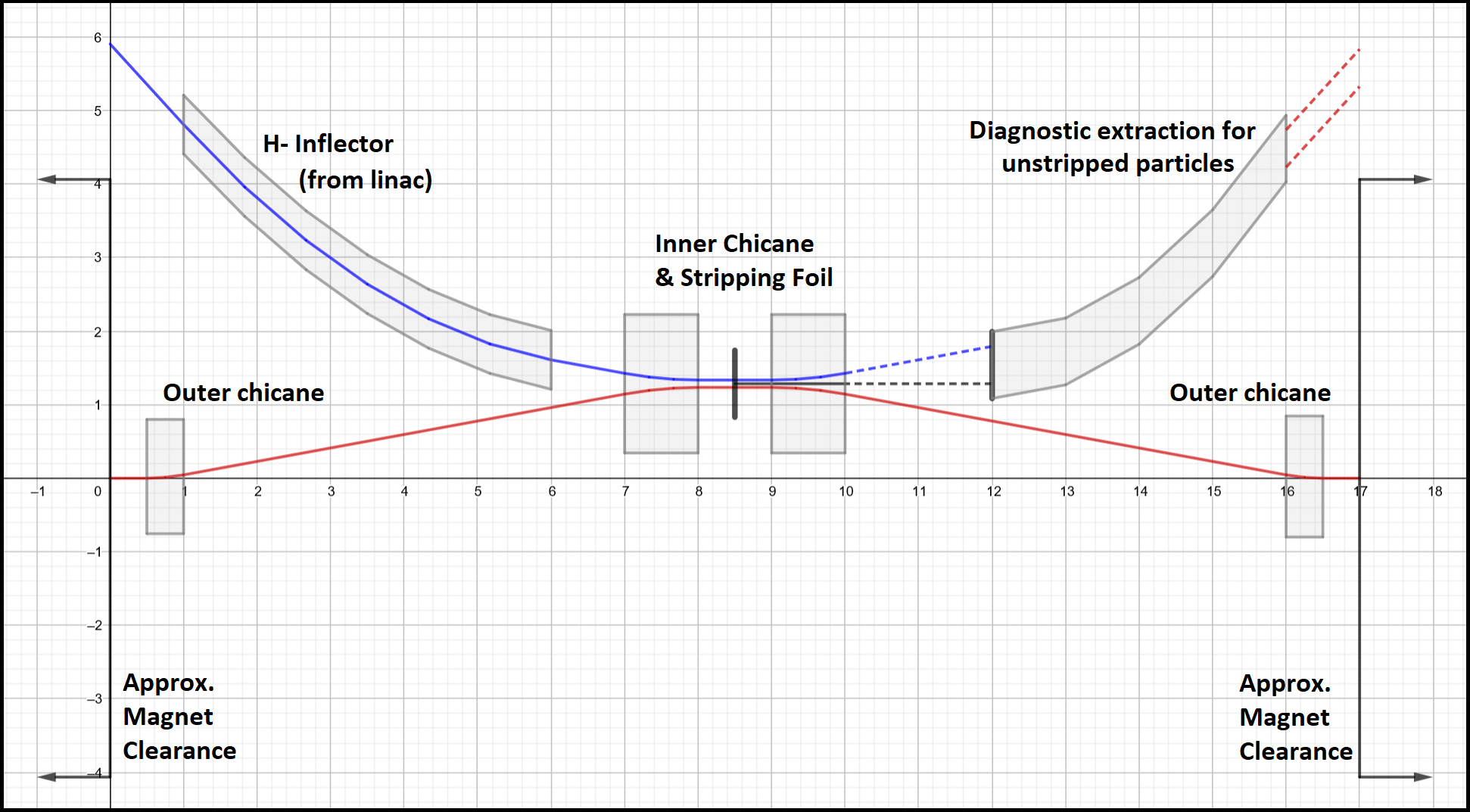}
 \caption{Example injection chicane for the foil-injection of a 2~GeV H$^{-}$ beam over a 17~m uninterrupted straight section. Beam travels from the left.  The vertical scale is magnified by 10 with respect to the horizontal scale (lengths in m).}
  \label{HminusChicane}
\end{centering}
\end{figure}

Fig.~\ref{HminusChicane} shows the foil injection scheme for a 2~GeV H$^{-}$ beam using a horizontal four-magnet chicane spanning 17~m. The inner chicane dipoles and the inflector magnets are limited to 0.17~T to keep the H$^{-}$ Lorentz stripping rate well below $10^{-6}$/m~\cite{Carneiro}. The outer chicane dipoles are at 0.34~T and can be powered on the same circuit as the inner chicane dipoles. The unstripped H$^{-}$ and H$^{0}$ particles can pass through a thicker second foil and extracted into a line for diagnostics and beam dump. In this preliminary design, the incoming linac beam clears the circulating beam orbit by 58.2 cm leaving adequate clearance for the horizontal width of the quadrupole focusing triplet. Injection painting magnets (not shown) may be placed in the 6~m between the inner and outer chicane magnets. The 1~meter separation between the inner chicane magnets leaves adequate space to improve the injection with new H$^{-}$ stripping methods, such as multiple thin foils or laser-stripping methods.


The injection design is very similar to that which has been successfully demonstrated at SNS~\cite{Raparia} and J-PARC~\cite{Hotchi}. In comparison to these facilities, the injection straight is longer because of the higher injection energy, but more compact than a simple scaling would indicate because the magnets that the injection line has to clear are smaller.

For this report, we have ensured our injection design is compatible with foil-stripping injection, a technique well-demonstrated in accelerator operations. We anticipate however, that the foil-stripping technique may ultimately be supplanted by the laser-stripping technology (currently demonstrated at the SNS over a 10~$\mu$s timescale with 95\% efficiency~\cite{Cousineau}). With a laser stripping upgrade of the AR scenario, the AR will no longer be limited by injection and would then be capable of supporting a MW-class 2~GeV pulsed program (see Section~\ref{sec:Science2}).

\subsection{Example RCS Lattice}  \label{sec:RCSlattice}

\begin{figure}[htp]
\begin{centering}
\includegraphics[height=200pt, width=280pt]{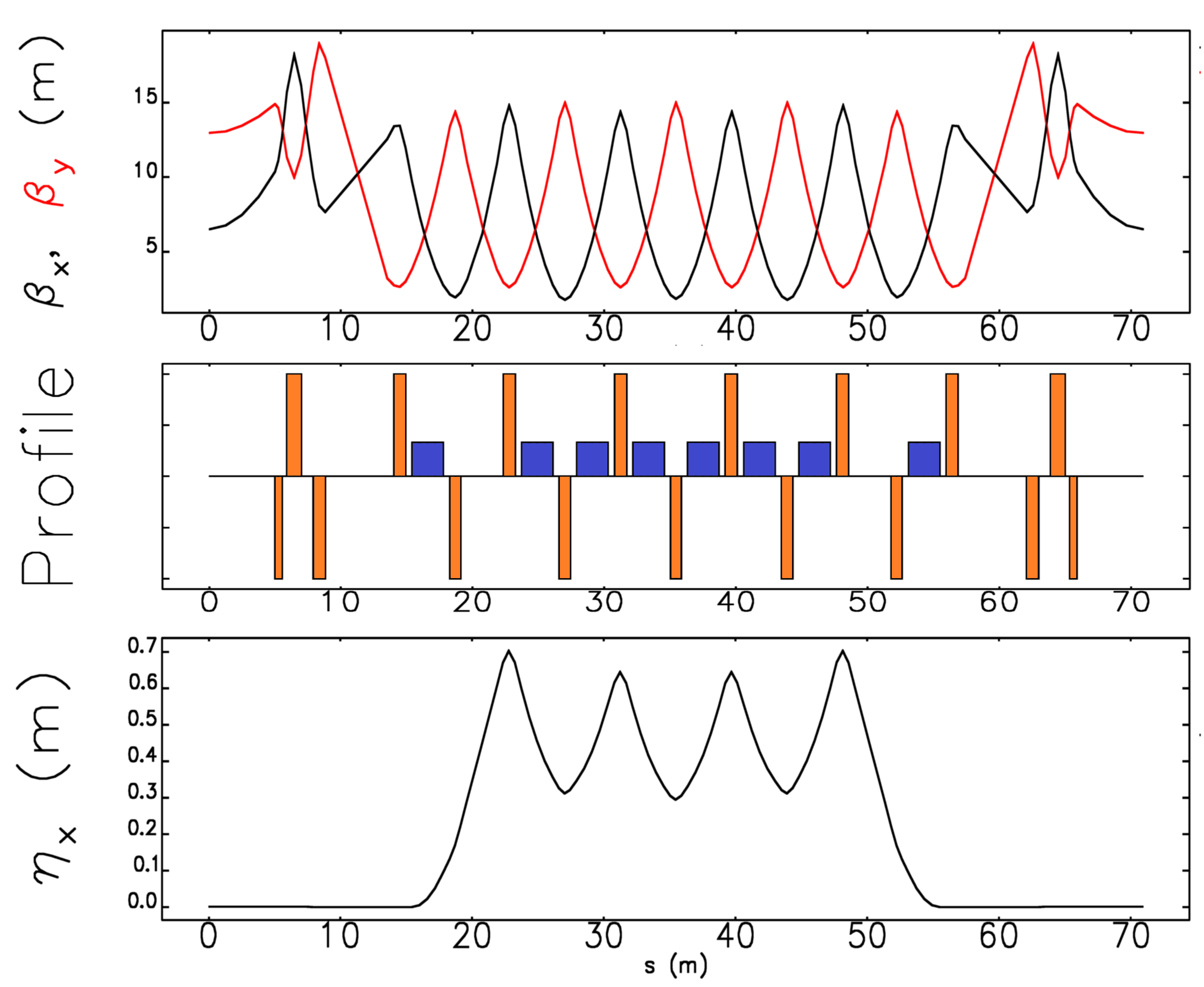}
 \caption{Twiss parameters for one of eight superperiods of the RCS lattice. (top) Horizontal and vertical beta functions are shown in black and red, respectively. (middle) Location and length of magnetic lattice elements with dipoles are shown as short blue rectangles and quadrupoles shown as tall orange rectangles. (bottom) Linear dispersion function.}
  \label{RCSlattice}
\end{centering}
\end{figure}

Many lattice solutions are compatible with the parameters outlined in Table~\ref{Param} and here we give an example RCS lattice in lieu of a full optimization study. Fig.~\ref{RCSlattice} shows the optics functions and magnet layout for one of eight superperiodic cells of an example RCS lattice. Table~\ref{LatticeParam} gives parameters specific to this particular RCS lattice design.
The lattice design is a series of achromatic FODO arcs each
 bookended by focusing triplets that convey the beam through long dispersion-free inserts. The insertion-region is 10 meters in order to accommodate a 2~GeV injection region. The insertion optics assume that the beam is transferred from a accumulator ring or that laser stripping injection is used; the beta functions and injection insertion length would be increased for a direct foil injection scenario.

In a FODO cell, the peak beta function is proportional to the cell length and consequently the beta functions and beam sizes  are minimized by the use of many alternating focusing quads. The tight-focusing combined with the ``missing dipole'' feature minimizes the momentum compaction factor to avoid transition crossing.

There is significant phase advance across the straight sections between bending arcs to allow for high-efficiency self-contained collimation units. Further, both the extraction kickers and the extraction septum can be placed between bending arcs to allow for clean beam separation without enlarged dipole apertures.

The lattice leaves space for dipole, tune, and chromaticity correctors (not shown). With the correctors, an operating point and dynamic aperture study would still be required to refine the lattice design.

\begin{table}[htp]
\centering
\caption{RCS and Injection Accumulator Ring lattice parameters.}
\begin{tabular}{| l || c | c | l |}
\hline
Lattice Parameter & RCS & AR & Unit \\
\hline
Circumference & 570 & 570 & m \\
Superperiodicity & 8 & 4 & \\
\hline
Maximum Beta functions ($x$,$y$) & 19,~19 & 32,~32 & m \\
Maximum Dispersion functions & 0.7 & 3.3 & m \\
Momentum compaction factor & 3.0 & 8.2 & $10^{-3}$ \\
\hline
Injection insertion length & 12 & 17 & m \\
Additional straights per superperiod & 2$~\times~$5.3 & 8$~\times~$5.5 & m \\
Number of dipoles per superperiod & 8 & 12 & \\
Number of quadrupoles per superperiod~ & 17 & 25 & \\
\hline
Dipole Field & 1.0 & 0.31 & T \\ 
Quadrupole Field & 14 & 4.2 & T/m \\
\hline
\end{tabular}
\label{LatticeParam}
\end{table}

At 2~GeV, a 95\% normalized emittance of 24 $\pi$ mm mrad and a maximum beta function of 30~m, the 8$\sigma$ beam size is 6.4~cm. Allowing for a further +/- 3~mm in orbit deviation, we find a pipe diameter of 7.0~cm or 2.25'' allows for a conservative margin for beam acceptance. If the collimator acceptance is set to 3$\sigma$ (with +/- 3~mm) the ratio between the collimator acceptance and pipe acceptance is 1.78, which is more conservative than the ratio of 1.5 used in J-PARC's precedent for the operation of MW-class beams~\cite{Hotchi}. The design allows for high efficiency transmission with well-controlled losses at the required $35\times 10^{12}$~protons, but also the possibility of considerably exceeding that intensity over time with careful optics tuning.


Table~\ref{LatticeParam} shows the required dipole and quadrupole fields for 8~GeV operation. At 8~GeV and 20~Hz, the maximum rate of change for the dipole magnetic field is 43~T/s, $\sim 25\%$ more rapid then the PIP-II Booster (35~T/s). The beampipe aperture and magnetic field strengths are quite comparable to the Main Injector, and would be considerably more affordable than magnets considered for large aperture facilities such as J-PARC, SNS, or the proposed ESS accumulator ring.

\begin{table}[bhtp]
\centering
\caption{RCS RF Parameters for 8~GeV, 20~Hz operation.}
\begin{tabular}{| l | c | l |}
\hline
Parameter & Value & Unit \\ \hline
RF Frequency Range & 50.326-52.812 & MHz  \\
Max RF Frequency Slew Rate & 248 & MHz/s \\
\hline
Total RF Voltage & 1.2 & MV  \\
Required num. RF cavities (at 60kV) & 20 & \\ 
Max Acc. Rate & 380 & GeV/s \\
\hline
Average Beam Current & 3.0 & A \\
\hline
\end{tabular}
\label{RCSRFParam}
\end{table}

As indicated in Table~\ref{RCSRFParam}, the required accelerating voltage for the RCS is 1.2~MV with a frequency range of 50.326-52.812~MHz. Existing 2.5~m (flange to flange) Booster RF cavities to be used for PIP-II operations accelerate protons at 60~kV~\cite{Tan}.  Perpendicular-biased RF cavities have been proposed to provide the same acceleration in half the longitudinal length~\cite{Hassan}. For either RF cavity design, 20-22 such RF cavities are required. The 20~m of dispersion-free straights in each superperiod can each accommodate up to eight parallel-biased RF cavities or sixteen perpendicular-biased RF cavities. Consequently using 60~kV parallel-biased RF cavities, only three superperiods are required, leaving enough space for injection, extraction, and collimation in the other three superperiods. This is a conservative estimate; with higher acceleration per unit length (such as a perpendicular bias cavity), considerable longitudinal real estate would be available for further upgrades to the RCS ramp rate. Conventional RF cavity R\&D is recommended to optimize the performance of a proposed RCS.

To minimize eddy-current heating effects, the beampipe inside of the RCS magnets should be constructed from ceramic, with brazed metal vacuum flanges on either side. On the other hand, the interior of the ceramic beampipe should use a thin metal layer to mitigate resistive wall instabilities~\cite{ICD2}. At 10~$\mu$m of Fe or Ni, the thin metal layer will be several skin-depths at frequencies above $\sim$100~MHz, while other frequencies (including synchrotron sidebands) can be mitigated by active dampers. The eddy current heating should be below $15~^{\circ}$C for ramp rates up to 30~Hz.

\subsection{\label{sec:ARlattice} Example 2 GeV Accumulator Ring Lattice}

\begin{figure}[htp]
\begin{centering}
\includegraphics[height=200pt, width=280pt]{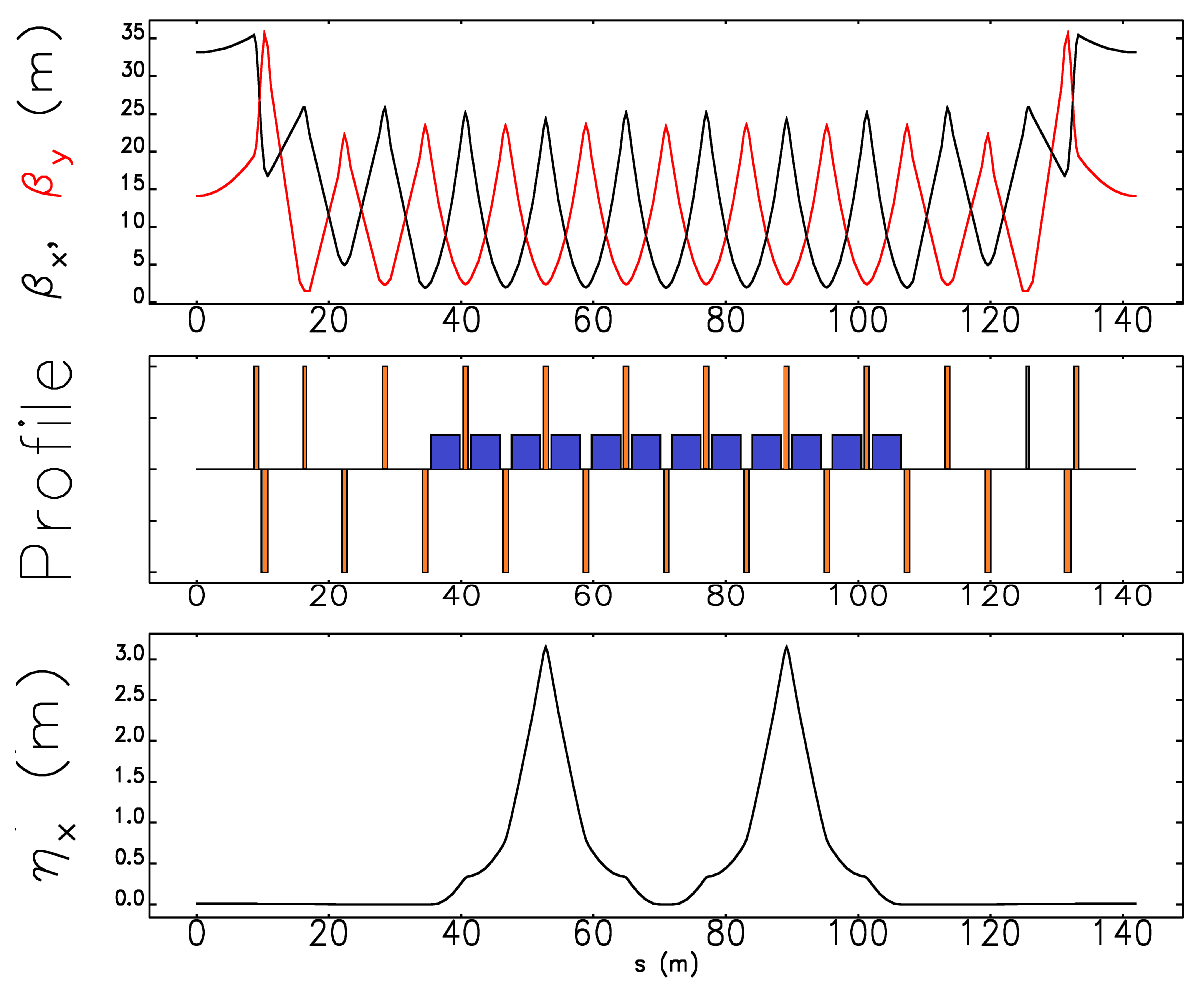}
 \caption{Twiss parameters for one of four superperiods of a 2 GeV Accumulator Ring lattice. (top) Horizontal and vertical beta functions shown in black and red, respectively. (middle) Location and length of magnetic lattice elements with dipoles shown as short blue rectangles and quadrupoles shown as tall orange rectangles. (bottom) Linear dispersion function.}
  \label{ARlattice}
\end{centering}
\end{figure}

For optimal performance of the RCS, a separate 2~GeV accumulator ring (AR) is envisioned which would receive beam from the linac, accumulate charge during the RCS cycle, and initialize proper bunching prior to transfer for RCS injection. Fig.~\ref{ARlattice} shows the optics functions and magnet layout for one of four superperiodic cells of an example AR lattice. Table~\ref{LatticeParam} gives parameters specific to this particular AR lattice design. The injection straight is 17~m, dispersion-free, and $\beta_{x} \times \beta_{y} = 440$ m$^{2}$ consistent with the injection chicane shown in Fig.~\ref{HminusChicane} and the foil-injection scenarios shown in Fig.~\ref{Foil}.

For the 2~GeV AR lattice, the maximum dipole field is 0.31~T and the maximum quadrupole field is 4.2~T/m. These field requirements are comparable to the CBETA permanent magnets~\cite{CBETA}. SmCo permanent magnets are recommended for their radiation resistance, which can withstand up to about $5\times 10^9$~rad before 1\% degradation~\cite{Blackmore}. Powered DC-magnets would likely be needed for the injection straight, to improve matching and to eliminate risk of demagnetization.

\subsection{\label{sec:RCSgreater} Greater Acceleration and Energy}

The RCS described above extracts 8 GeV protons and operates at 20~Hz. However, the same RCS lattice design can accommodate sufficient RF for 8 GeV at 30~Hz or 12~GeV at 20~Hz. Table~\ref{Ramp12} compares the ramp parameters for these three scenarios. For the 12~GeV scenario the peak dipole field would be 1.4~T, which is still a reasonable maximum dipole field in line with the maximum field of the Main Injector dipole.

To meet the requirements of the 2.4~MW DUNE/LBNF program at reduced ramp rate, the 30~Hz scenario only needs $33.5\times 10^{12}$ protons, compared to $35\times 10^{12}$ protons at 20~Hz. The primary beneficiary would be the 8 GeV experimental program from the RCS beamline.

For the 12~GeV scenario, the primary consideration should be the Main Injector program. At 12 GeV Main Injector injection, neither the Main Injector space-charge tune-spread nor the geometric emittance will exceed that of PIP-II era Main Injector operation. However, in that case the MI-8 and RR injection cannot be re-used and instead the RCS would require a new transfer line directly into the MI (on Fig.~\ref{fig:upgradeOpts} middle, as opposed to left). Consequently, a 12~GeV RCS program cannot be easily retrofit onto an 8~GeV program using MI-10 injection from the start. 

For the 20~Hz 12~GeV scenario the beam power available at 12~GeV concurrent with the 120~GeV LBNF program would be comparable to 8~GeV beam power available from the 30~Hz 8~GeV scenario, although some experiments may specifically prefer the higher extraction energy.

\begin{table}[htp]
\centering
\caption{RCS Ramp Parameters.}
\begin{tabular}{| l | c | c | c | l |}
\hline
   & 8~GeV, 20~Hz & 8~GeV, 30~Hz & 12~GeV. 20~Hz &  \\
\hline
Min. RCS Intensity for 2.4~MW & 35 & 33.5 & 35 & $10^{12}$ \\
Available RCS Power* & 0.75 & 1.2 & 1.1 & MW \\
\hline
RF Frequency Range & 50.326-52.812 & 50.326-52.812 & 50.326-52.965 & MHz \\
Max RF Frequency Slew Rate & 248  & 372 & 325 & MHz/s \\
\hline
Total RF Voltage & 1.2 & 1.9 & 2.1 & MV \\
Required num. RF cavities (at 60kV) & 20 & 32 & 35 & \\ 
Max Acc. Rate & 381 & 572 & 633 & GeV/s \\
\hline
Max Dipole Field & 1.0 & 1.0 & 1.4 & T \\
Max Dipole Slew Rate & 43 & 65 & 49 & T/s \\
Max Quad Field & 14 & 14 & 20 & T/m \\
\hline
\end{tabular}
*concurrent with 120 GeV MI operations
\label{Ramp12}
\end{table}

Higher than 30~Hz ramp rate may also be possible for this RCS design. Firstly, it would require RF cavities with a higher acceleration gradient per unit length then existing Booster RF cavities, such as the perpendicular-biased RF cavity design. Secondly, the ceramic beampipe will need to be complex, with RF-shielding in longitudinal stripes similar to the J-PARC design~\cite{Kinsho}.

\section{\label{subsec:mi} Main Injector Operations}

\subsection{Beam Transfer and Injection} 
The circumference and RF extraction frequency of the RCS are expected to be a sub-harmonic of the MI (h=588), such that the 8 GeV injection directly into the Main Injector is in a synchronous bunch-to-bucket box car fashion. The location of the RCS will be such that its extracted beam transport line will connect to the existing 8 GeV tunnel, which approaches the MI tunnel at the MI-10 location. The MI-10 location currently houses direct injection into the Main Injector from the Booster, direct injection into the Recycler from the Booster and the horizontally bending switch magnet to send beam to the 8-GeV neutrino experimental area. In the LBNF era, the Main Injector MI-10 injection straight section will be decommissioned and converted into an extraction area for the new DUNE project. 

Since the MI-10 straight section is not available for injection from the new RCS, the other potential options include the MI-22 straight section (previously used for proton extraction from Recycler) and the MI-30 straight section (currently used for Recycler to Main Injector injection). 

We assume that the Recycler ring as an accumulation ring will be abandoned. The current 8 GeV line to Recycler trajectory can be maintained and the section of the Recycler between MI-10 and either MI-22 or MI-30 can be maintained as a transfer line. 

\begin{figure}[htp]
\begin{centering}
\includegraphics[height=200pt, width=400pt]{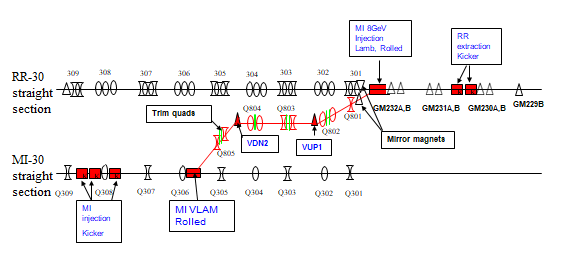}
 \caption{Schematic view of the current Recycler to MI transfer line in MI-30 straight}
  \label{RR_to_MI}
\end{centering}
\end{figure}

The standard technique for injection will consist of a Lambertson and kicker separated by 90 degrees in betatron phase advance. The orientation of the bend field of the injection kicker is dependent on the available locations for the kicker. The Lambertson bend region will be in the opposite plane. For example at MI-10 the Recycler injection kicker is in the horizontal direction and is located just upstream of the horizontal-focusing quad Q104, with the Lambertson located just upstream of Q102 having a vertical bend field. On the other hand, injection into the MI utilized a horizontal bend field in the Lambertson and a vertical kicker. The current injection from the Recycler to the Main Injector uses a vertical Lambertson at Q306 and a horizontal kicker at Q308.
The existing transfer line between the Recycler and the Main Injector currently in use is shown in Figure ~\ref{RR_to_MI}.

Very few modifications to this transport line will be required. Since the Recycler is no longer a ring, the kicker and Lambertson in the Recycler ring are no longer needed.

If we select MI-22 as the injection point from the RCS transport line, it will require the use of a vertical kicker located just upstream of Q223. The utilization of a Lambertson just upstream of Q221 may be problematic due to the orientation of the Main Injector ring to the outside radius of the tunnel wall. In this case a vertical C-magnet (or DC magnetic septum) will be considered.

\subsection{RF Considerations}
\begin{figure}[htp]
\begin{centering}
\includegraphics[height=200pt, width=280pt]{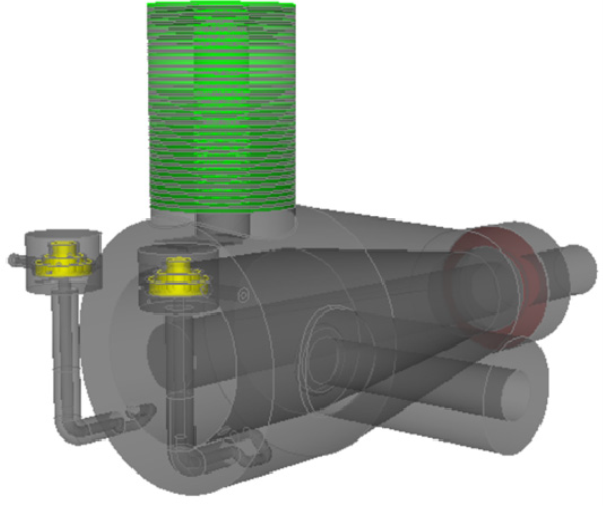}
 \caption{New MI Cavity conceptual design.}
  \label{MICavity}
\end{centering}
\end{figure}

The current MI RF system with two power amplifiers per station has enough power to accelerate up to 120$\times 10^{12}$ protons at 240 GeV/sec. A new MI RF system will be needed to accelerate the higher beam intensities required for 2.4 ~MW. The high level specifications of the new MI RF system are outlined in Table~\ref{RFSpecs}.  The frequency range has increased to 490 KHz to accommodate injection at 6 GeV. The maximum acceleration rate is 240 GeV/sec and, given the revolution period of Main Injector, requires 2.7 MV for acceleration. The maximum required voltage is determined by the bucket area, allowing enough overhead to operate with up to two RF cavities down. The peak RF power is 7.1 MW and the average is 3.6 MW, since the RF is off for half the Main Injector Cycle. The average beam current is 2.7 A while the fundamental RF current can be as high as 5.2 A after transition. The new 53 MHz cavities will be powered by the Eimac 8973 power tetrode with a power capability greater than 1~MW, operating frequency up to 110~ MHz and a plate dissipation of 1,000~kW.
Figure ~\ref{MICavity} shows the new MI cavity conceptual design including HOM dampers with high-pass filters and the input coupler. 

\begin{table}[htp]
\centering
\caption{RF System Specifications}
\begin{tabular}{|| l || c || l ||}
\hline
Parameter & Value & Unit \\
\hline
Frequency & 52.617-53.104  & MHz \\
Max. Acc. Rate &  240 & GeV/s \\
Frequency Slew Rate &  1.6 & MHz/sec \\
Acceleration Voltage &  2.7 & MV \\
Peak Beam Power  &  7.1 & MW  \\
Average Beam Power & 3.6 & MW \\
Peak Voltage  & 4.8  & MV \\
Average Beam Current & 2.7 & A \\
Fundamental RF Current & 4.6-5.2 & A \\
\hline
\end{tabular}
\label{RFSpecs}
\end{table}

\subsection{Space Charge Effects and Beam Instabilities}
\begin{figure}[htp]
\begin{centering}
\includegraphics[height=200pt, width=280pt]{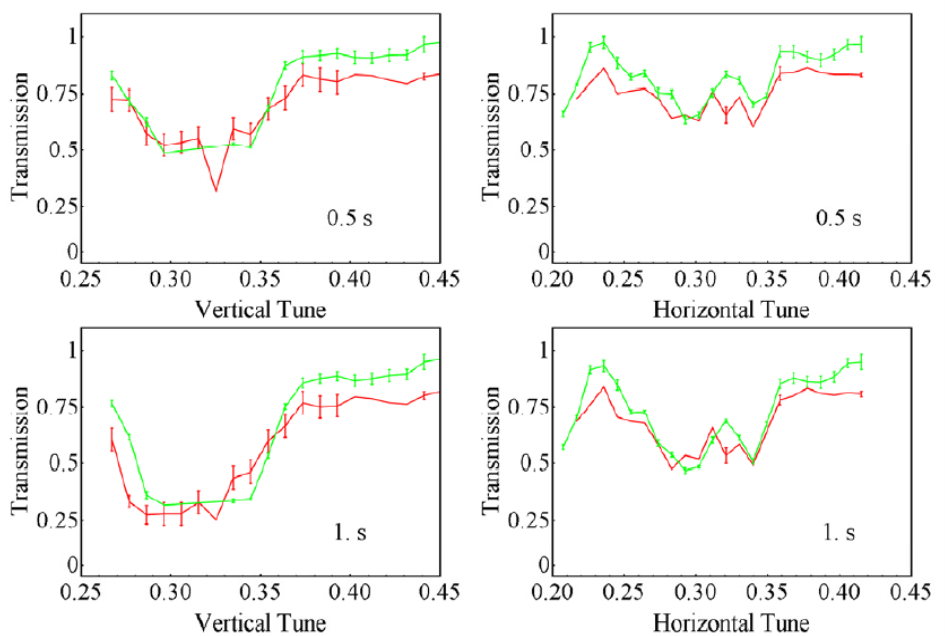}
 \caption{Transmission efficiency for 2 bunches with different intensities and bunch lengths.  Red trace: $182\times 10^9$, $\sigma_{b}$=4 nsec. Green trace: $55\times 10^9$, $\sigma_{b}$=1.2 nsec}
  \label{EfficiencyBF}
\end{centering}
\end{figure}

\begin{figure}[htp]
\begin{centering}

\includegraphics[height=180pt]{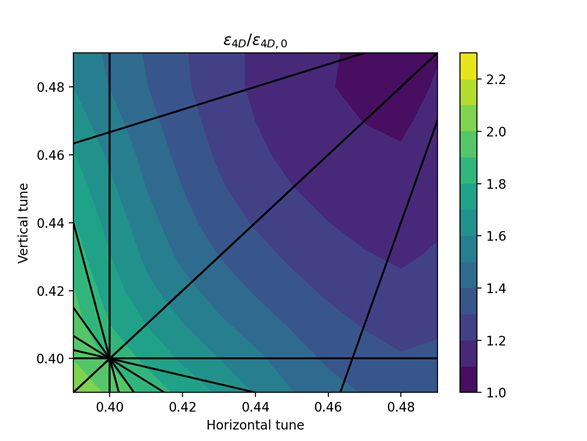}
\includegraphics[height=180pt]{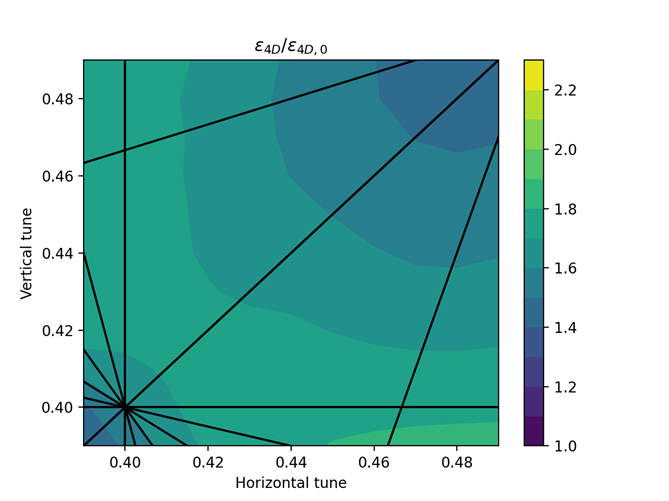}
 \caption{Emittance growth (4D) in MI for $2\times 10^{11}$ protons per bunch (top) and $4\times 10^{11}$ protons per bunch (bottom).} 
  \label{4DemittanceINT}
\end{centering}
\end{figure}

\begin{figure}[htp]
\begin{centering}

\includegraphics[height=180pt]{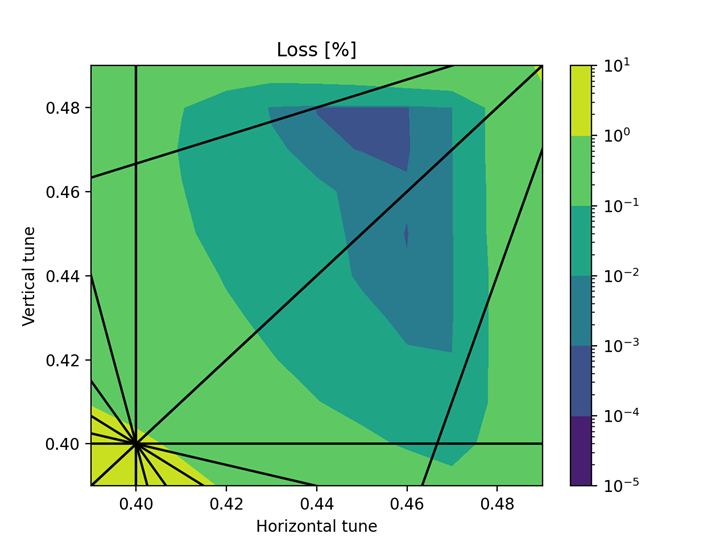}
\includegraphics[height=180pt]{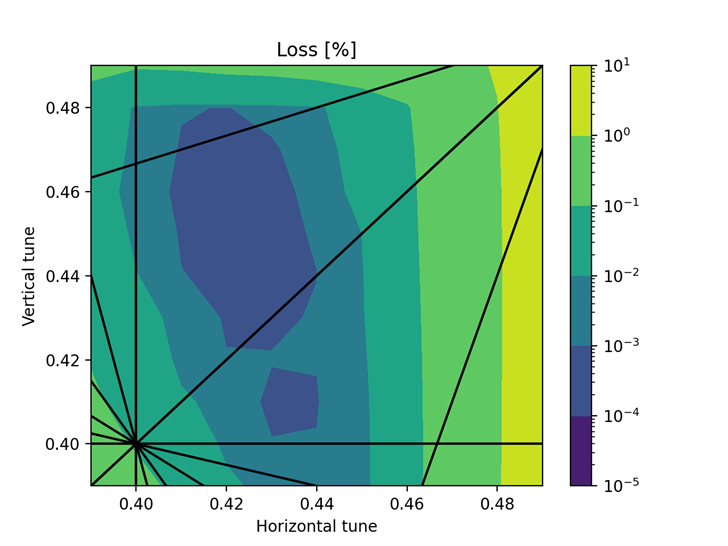}
\includegraphics[height=180pt]{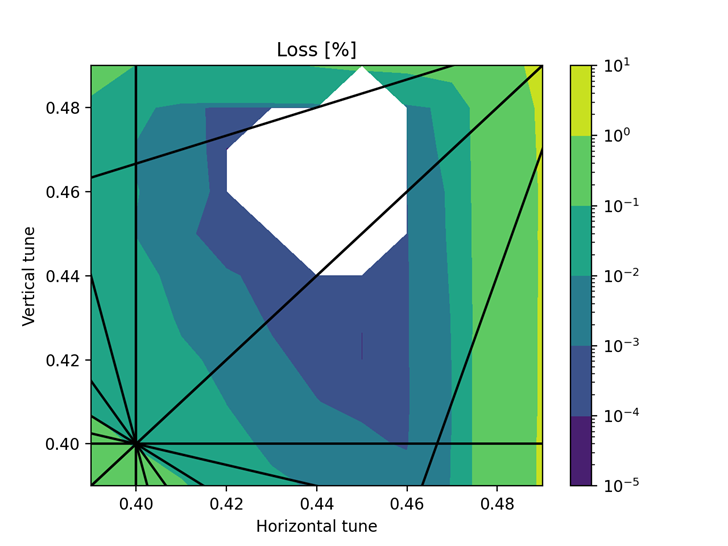}
 \caption{Loss plots for $4\times 10^{11}$ protons per bunch in the MI.  (top) Bunch length of $\sigma$=0.5~m at 8 GeV.  (center) Bunch length of $\sigma$=0.8~m at 8 GeV. (bottom) Bunch length of $\sigma$=0.5~m at 10 GeV. The white area in tune space indicates a no-loss region.} 
  \label{LossBL}
\end{centering}
\end{figure}

Space charge effects can lead to transverse emittance growth resulting in losses. The effect of the bunching factor on beam transmission efficiency has been measured in MI. The transmission efficiency as a function of tune for two different bunch intensities with the same transverse emittance, though with different bunch lengths, is shown in Figure~\ref{EfficiencyBF}. The code Synergia3 was used to simulate 2D scans in betatron tunes (0.39-0.49) and evaluate emittance growth in the MI for different beam intensities. The 4D emittance (product of the horizontal and vertical emittances) growth for two different bunch intensities is shown in Figure~\ref{4DemittanceINT}.   The effect of bunch length on losses for $4\times 10^{11}$ protons per bunch is shown in Figure~\ref{LossBL}. Furthermore, simulations have been performed using Synergia3 to compare the effect of increasing intensities in the Main Injector for different injection energies ranging from 8~GeV to 10~GeV.  A plot of the losses during the tune scans for two different MI injection energies is shown in Figure~\ref{LossBL}.

Instabilities can lead to losses and, if severe enough, may require the beam to be sent to the abort before it can be extracted. Coupled bunch instabilities are expected to be controlled by the transverse and longitudinal damper already installed. Single bunch instabilities are to be investigated further using the collective effects code Synergia3. 



\subsection{Transition Crossing}
\begin{figure}[htp]
\begin{centering}
\includegraphics[height=200pt, width=280pt]{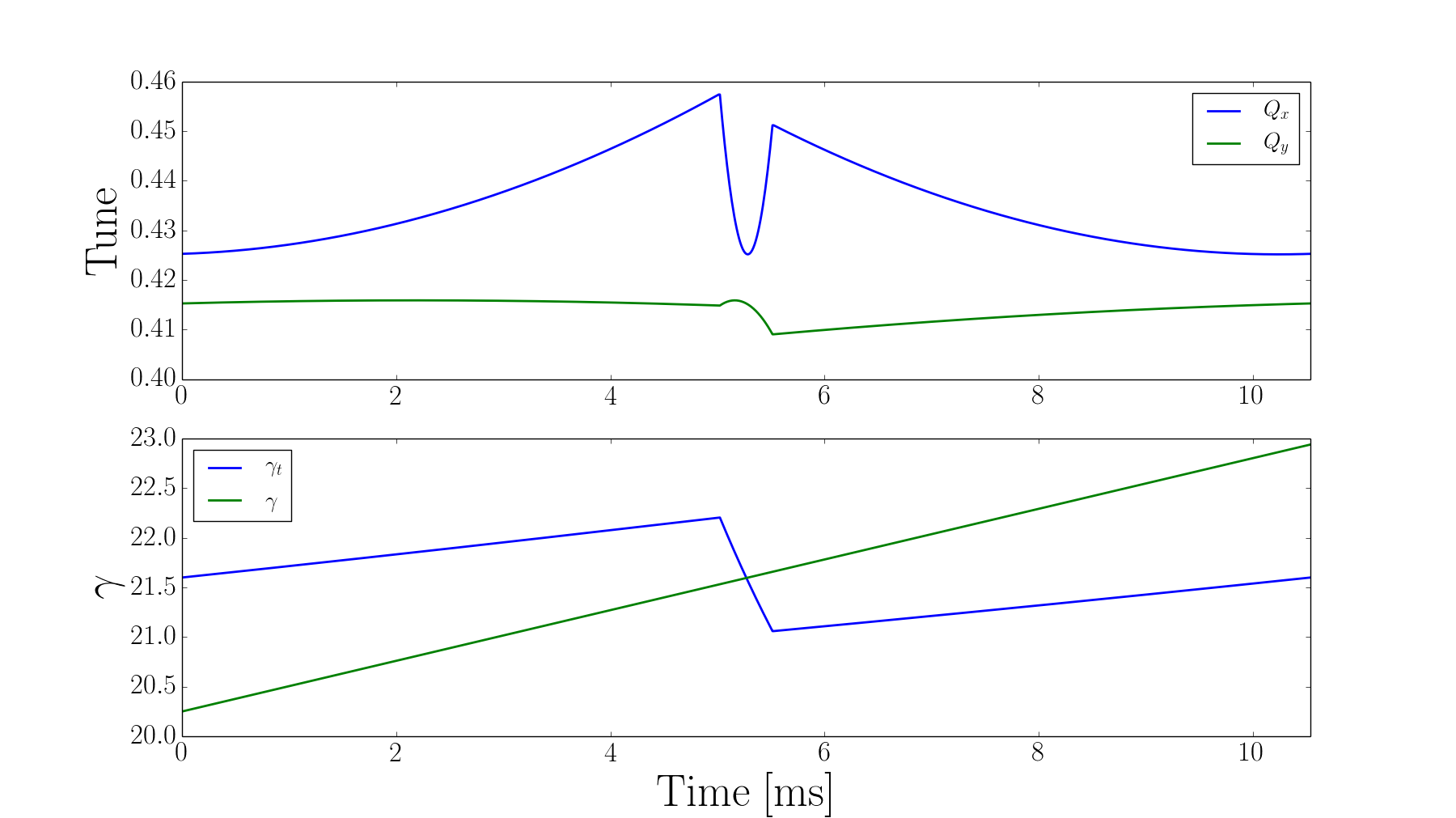}
 \caption{The effect on the tunes and on the $\gamma_{t}$ of the lattice during the jump scheme.}
  \label{Jump}
\end{centering}
\end{figure}

\begin{figure}[htp]
\begin{centering}

\includegraphics[height=180pt]{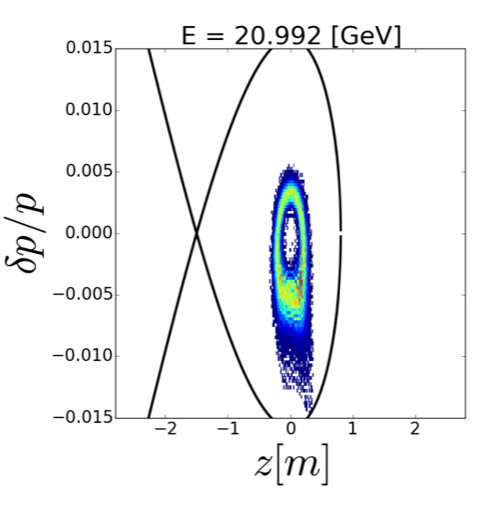}
\includegraphics[height=180pt]{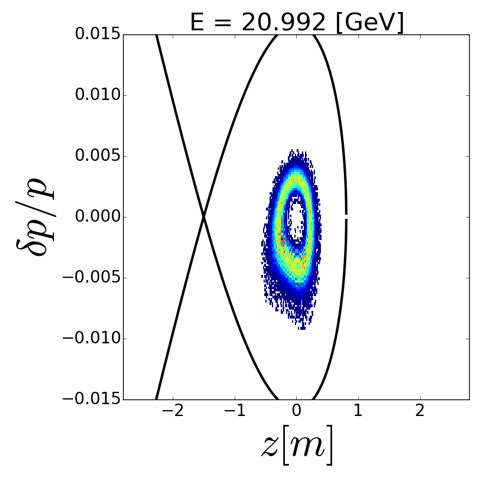}
 \caption{The longitudinal phase space distribution after transition of PIP-II beam without (left) and with the $\gamma_{t}$ jump (right).} 
  \label{TransitionPS}
\end{centering}
\end{figure}

Transition crossing in the MI can result in longitudinal emittance blow-up, and instabilities that can create unacceptable losses. For PIP II a first order $\gamma_{t}$ jump has been developed and will be installed, eliminating transition losses and keeping longitudinal emittance under control. The  $\gamma_{t}$ jump scheme maintains a minimum distance from the transition energy and crosses transition almost 10 times faster. The jump consists of 4 sets of pulsed triplets. Each triplet has two quads in the arc and one of twice the integrated strength in the dispersion free straight section. 
The transition crossing in the MI under PIP II conditions was simulated using Synergia taking into effect not only the longitudinal dynamics but also the transverse effects from the lattice optics changes. Figure~\ref{Jump} shows the effect on $\gamma_{t}$ and on the tunes during the jump. The effect of the $\gamma_{t}$ jump on the longitudinal phase space after transition is shown in Figure~\ref{TransitionPS}.
The simulations did not include longitudinal space charge and Z/n since space charge effects were not expected to have a large effect due to the large longitudinal emittance. In the case of our upgrade scenario where no slip stacking is used, the longitudinal emittance is expected to be much smaller and the space charge effects during transition crossing need to studied.

\subsection{Cycle Time Improvements} \label{sec:4MW}

\begin{figure}[htp]
\begin{centering}
\includegraphics[height=200pt, width=280pt]{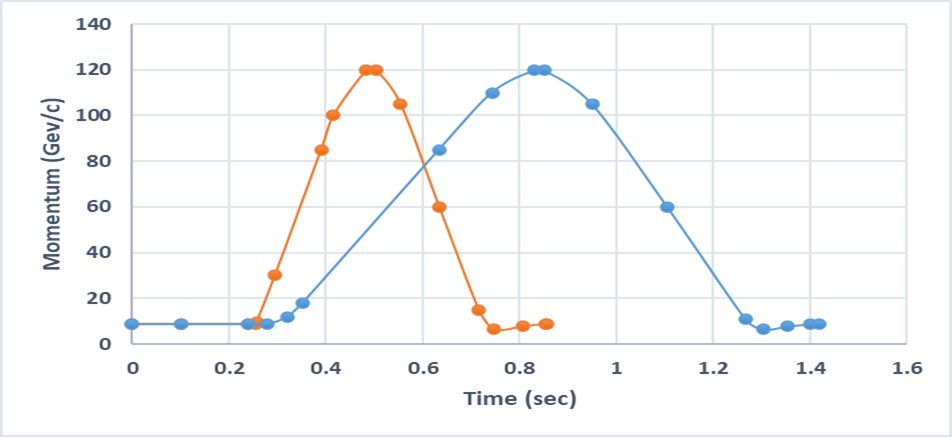}
 \caption{Comparison between the regular MI Ramp and the Ramp with 600 GeV/sec.}
  \label{MIRamps}
\end{centering}
\end{figure}

\begin{figure}[htp]
\begin{centering}
\includegraphics[height=200pt, width=280pt]{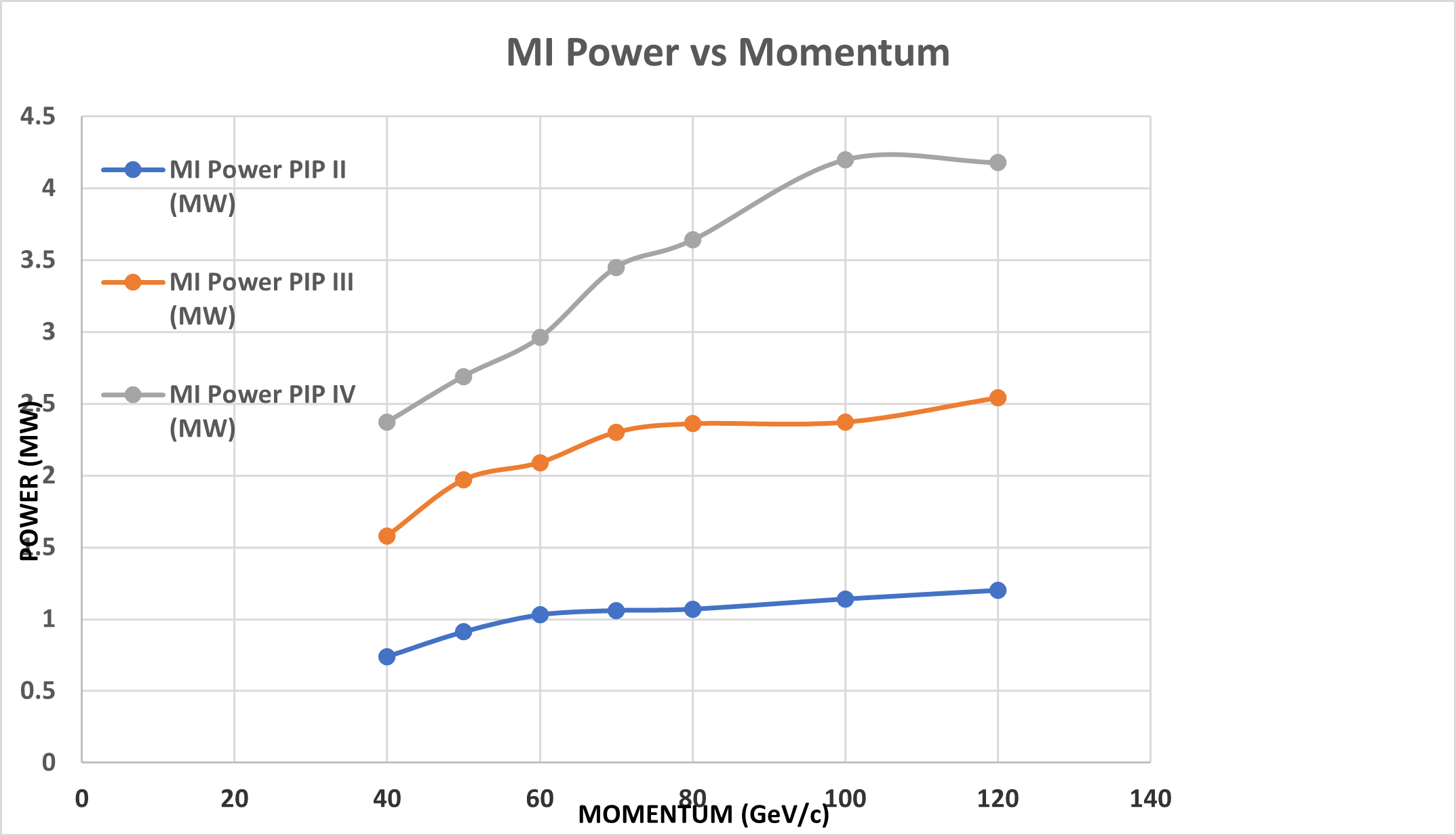}
 \caption{MI Power vs Momentum for the different Upgrade Stages.}
  \label{MIPower}
\end{centering}
\end{figure}

Since it is difficult to further increase the beam power by increasing the beam intensity we need to look at ways to reduce the MI cycle time.  In the absence of the Recycler for our scenario, the MI will require 5 injections from the RCS to fill, and assuming a 20~Hz RCS rep rate, then the total cycle time required for the MI to fill and accelerate to 120 GeV is 1.4 sec.  Increasing the acceleration rate in the MI by a factor of 2.5 from 240 GeV/sec to 600 GeV/sec will reduce the MI cycle time from 1.4 sec to 0.9 sec. This will increase the beam power at 120 GeV from 2.5 MW to 4.2 MW. The comparison between the two MI ramps is shown in Figure~\ref{MIRamps}. The MI power vs momentum for the different stages is shown in Figure~\ref{MIPower}. Increasing the acceleration rate will require more power supply voltage and more RF voltage and power.

In order to increase the Power Supply voltage, we will need to add 2 dipole power supplies and 1 Quad power supply in every MI service Building. The service buildings will have to be enlarged to accommodate the extra power supplies. Power supply transformers, pads and additional feeders need to be added outside each building. Two additional transformers and harmonic filters will also be needed in the power substation. The substation building will have to be expanded to accommodate additional breakers and equipment.

The new MI RF stations will have the power required to accelerate $1.85\times 10^{14}$ protons at 600 GeV/sec but additional RF voltage will be required. The minimum acceleration voltage for 600 GeV/sec is 6.75 MV and, assuming a maximum bucket area of 0.65 eV-sec after transition, a total of 7.5 MV of RF voltage will be required. Assuming 240 kV per cavity to achieve the required voltage will require at least 31 RF stations. Since the new cavity design is shorter than the current MI cavity we should be able to fit up to 33 RF cavities in the MI-60 straight section used for the RF.

\newpage

\section{\label{subsec:summary2} Accelerator Summary}

The Fermilab accelerator complex has driven decades of discoveries in accelerator-based science and comprises one of the world’s most productive scientific facilities. The performance that has been achieved in each accelerator is far beyond the original design specifications, and with FNAL’s transition to the intensity frontier over the last decade, the complex necessarily operates in regimes that were not previously envisioned.  As Fermilab moves steadily into the PIP-II and LBNF/DUNE era, key elements of the complex are reaching their fundamental limitations and must be upgraded or replaced.  Clearly, for any given upgrade scenario, a full campaign of planning exercises, simulations and analyses will be necessary.

As the fundamental bottleneck toward higher power, replacement of the Booster synchrotron is common to all upgrade scenarios.  A core element of the upgrade path presented here is a rapid cycling synchrotron that has been optimized not only for the high intensities required by LBNF/DUNE but also for compatibility with future upgrades based on evolving R\&D in Accelerator Science and Technology.

Targeted upgrades to the PIP-II linac and its supporting infrastructure are also a key component of this upgrade path.  Increasing the linac energy to 2 GeV would significantly reduce space-charge effects at RCS injection.  The PIP-II linac tunnel can be extended by approximately 400 ft, providing adequate room for the delivery of the energy upgrade without requiring a new environmental assessment; furthermore, the energy upgrade can be progressive, and a pulsed-only mode of operation can be considered to significantly reduce the associated costs.  

Increasing the linac current would help alleviate challenges associated with foil injection into the RCS.  A moderate upgrade to 4-5 mA is feasible without significant design changes and only requires a commensurate increase in the output power of the RF amplifiers;  however, moving beyond 5 mA would require the extensive, systemic evaluations and upgrades that are detailed in section III.B.5. Alternatively, a companion fixed-energy (2 GeV) accumulator ring is also considered as a future upgrade.  This accumulator ring would improve injection into the RCS while also supporting a growing portfolio of low-energy science opportunities.

In this upgrade scenario, with the higher intensity and repetition rate of the RCS obviating the need for slip stacking, the Recycler Ring would likely be retired.  If the peak energy of the RCS remains at 8 GeV, then existing sections of the RR can be maintained as a transfer line to the Main Injector.
Future upgrades to the MI power systems would reduce the MI cycle time and enable significantly higher power delivery to LBNF/DUNE without an increase in peak intensity.  This will require conventional RF-cavity R\&D, as well as significant conventional facilities and infrastructure upgrades, but provides a relatively straightforward path to accelerator-complex performance beyond 2.5 MW.

Throughout this planning exercise we have sought only to define the broad technological and strategic contours for various upgrade paths.  Generally, all considerations have been subordinate to the rapid, effective delivery of the requisite performance for the high-energy neutrino program; however, a significant alteration of the accelerator complex presents a unique opportunity in which, with careful forethought, the Lab can be positioned to capture a broad portfolio of science opportunities that complement the DUNE program, both scientifically and operationally.  In this upgrade, Fermilab can also lay the foundation for accelerator performance beyond the LBNF/DUNE era, and where possible, compatibility with the operationalized products of R\&D in Accelerator Science and Technology should be emphasized.

\newpage



\bibliography{FNAL_Upgrade}

\providecommand{\noopsort}[1]{}\providecommand{\singleletter}[1]{#1}%
\begin{thebibliography}{35}%
\makeatletter
\providecommand \@ifxundefined [1]{%
 \@ifx{#1\undefined}
}%
\providecommand \@ifnum [1]{%
 \ifnum #1\expandafter \@firstoftwo
 \else \expandafter \@secondoftwo
 \fi
}%
\providecommand \@ifx [1]{%
 \ifx #1\expandafter \@firstoftwo
 \else \expandafter \@secondoftwo
 \fi
}%
\providecommand \natexlab [1]{#1}%
\providecommand \enquote  [1]{``#1''}%
\providecommand \bibnamefont  [1]{#1}%
\providecommand \bibfnamefont [1]{#1}%
\providecommand \citenamefont [1]{#1}%
\providecommand \href@noop [0]{\@secondoftwo}%
\providecommand \href [0]{\begingroup \@sanitize@url \@href}%
\providecommand \@href[1]{\@@startlink{#1}\@@href}%
\providecommand \@@href[1]{\endgroup#1\@@endlink}%
\providecommand \@sanitize@url [0]{\catcode `\\12\catcode `\$12\catcode
  `\&12\catcode `\#12\catcode `\^12\catcode `\_12\catcode `\%12\relax}%
\providecommand \@@startlink[1]{}%
\providecommand \@@endlink[0]{}%
\providecommand \url  [0]{\begingroup\@sanitize@url \@url }%
\providecommand \@url [1]{\endgroup\@href {#1}{\urlprefix }}%
\providecommand \urlprefix  [0]{URL }%
\providecommand \Eprint [0]{\href }%
\providecommand \doibase [0]{https://doi.org/}%
\providecommand \selectlanguage [0]{\@gobble}%
\providecommand \bibinfo  [0]{\@secondoftwo}%
\providecommand \bibfield  [0]{\@secondoftwo}%
\providecommand \translation [1]{[#1]}%
\providecommand \BibitemOpen [0]{}%
\providecommand \bibitemStop [0]{}%
\providecommand \bibitemNoStop [0]{.\EOS\space}%
\providecommand \EOS [0]{\spacefactor3000\relax}%
\providecommand \BibitemShut  [1]{\csname bibitem#1\endcsname}%
\let\auto@bib@innerbib\@empty
\bibitem [{\citenamefont {Eldred}\ \emph {et~al.}(2019)\citenamefont {Eldred},
  \citenamefont {Lebedev},\ and\ \citenamefont {Valishev}}]{Eldred}%
  \BibitemOpen
  \bibfield  {author} {\bibinfo {author} {\bibfnamefont {J.}~\bibnamefont
  {Eldred}}, \bibinfo {author} {\bibfnamefont {V.}~\bibnamefont {Lebedev}},\
  and\ \bibinfo {author} {\bibfnamefont {A.}~\bibnamefont {Valishev}},\
  }\bibfield  {title} {\bibinfo {title} {{R}apid-{C}ycling {S}ynchrotron for
  {M}ulti-{M}egawatt {P}roton {F}acility at {F}ermilab},\ }\href
  {https://arxiv.org/abs/1903.12408} {\bibfield  {journal} {\bibinfo  {journal}
  {Journal of Instrumentation}\ }\textbf {\bibinfo {volume} {14}}\bibinfo
  {number} { (P07021)}}\BibitemShut {NoStop}%
\bibitem [{\citenamefont {Abi}\ \emph {et~al.}(2020)\citenamefont {Abi} \emph
  {et~al.}}]{DUNE}%
  \BibitemOpen
\bibfield  {number} {  }\bibfield  {author} {\bibinfo {author} {\bibfnamefont
  {B.}~\bibnamefont {Abi}} \emph {et~al.} (\bibinfo {collaboration} {DUNE
  collaboration}),\ }\bibfield  {title} {\bibinfo {title} {{Deep Underground
  Neutrino Experiment (DUNE), Far Detector Technical Design Report, Volume II:
  DUNE Physics}},\ }\href {https://arxiv.org/abs/2002.03005} {\bibfield
  {journal} {\bibinfo  {journal} {FERMILAB-PUB-20-025-ND}\ } (\bibinfo {year}
  {2020})}\BibitemShut {NoStop}%
\bibitem [{\citenamefont {Lebedev}\ \emph {et~al.}(2018)\citenamefont {Lebedev}
  \emph {et~al.}}]{PIP2}%
  \BibitemOpen
  \bibfield  {author} {\bibinfo {author} {\bibfnamefont {V.}~\bibnamefont
  {Lebedev}} \emph {et~al.},\ }\bibfield  {title} {\bibinfo {title} {{T}he
  {PIP}-{II} {C}onceptual {D}esign {R}eport {V}.0.3},\ }\href
  {https://pxie.fnal.gov/PIP-II_CDR/PIP-II_CDR_v.0.3.pdf} {\bibfield  {journal}
  {\bibinfo  {journal} {FERMILAB-TM-2649-AD-AIPC}\ } (\bibinfo {year}
  {2018})}\BibitemShut {NoStop}%
\bibitem [{\citenamefont {Cao}\ \emph {et~al.}(2017)\citenamefont {Cao} \emph
  {et~al.}}]{ICFA}%
  \BibitemOpen
  \bibfield  {author} {\bibinfo {author} {\bibfnamefont {J.}~\bibnamefont
  {Cao}} \emph {et~al.} (\bibinfo {collaboration} {ICFA Neutrino Panel}),\
  }\bibfield  {title} {\bibinfo {title} {{Roadmap for the international,
  accelerator-based neutrino programme}},\ }\href
  {http://icfa.fnal.gov/wp-content/uploads/2016-05-07-nuPanel-roadmap-Final.pdf}
  {\bibfield  {journal} {\bibinfo  {journal} {FERMILAB-FN-1031}\ } (\bibinfo
  {year} {2017})}\BibitemShut {NoStop}%
\bibitem [{\citenamefont {Abe}\ \emph {et~al.}(2018)\citenamefont {Abe} \emph
  {et~al.}}]{T2K}%
  \BibitemOpen
  \bibfield  {author} {\bibinfo {author} {\bibfnamefont {K.}~\bibnamefont
  {Abe}} \emph {et~al.} (\bibinfo {collaboration} {Hyper-Kamiokande
  Proto-Collaboration}),\ }\bibfield  {title} {\bibinfo {title} {{Physics
  potentials with the second Hyper-Kamiokande detector in Korea}},\ }\href
  {https://academic.oup.com/ptep/article/2018/6/063C01/5041972} {\bibfield
  {journal} {\bibinfo  {journal} {PTEP}\ }\textbf {\bibinfo {volume} {6}}
  (\bibinfo {year} {2018})}\BibitemShut {NoStop}%
\bibitem [{\citenamefont {Chakraborty}\ \emph {et~al.}(2018)\citenamefont
  {Chakraborty}, \citenamefont {Deepthia},\ and\ \citenamefont
  {Goswamia}}]{ESS}%
  \BibitemOpen
  \bibfield  {author} {\bibinfo {author} {\bibfnamefont {K.}~\bibnamefont
  {Chakraborty}}, \bibinfo {author} {\bibfnamefont {K.~N.}\ \bibnamefont
  {Deepthia}},\ and\ \bibinfo {author} {\bibfnamefont {S.}~\bibnamefont
  {Goswamia}},\ }\bibfield  {title} {\bibinfo {title} {{Spotlighting the
  sensitivities of Hyper-Kamiokande, DUNE and ESS$\nu$SB}},\ }\href
  {https://www.sciencedirect.com/science/article/pii/S0550321318302943}
  {\bibfield  {journal} {\bibinfo  {journal} {Nuc. Phys. B}\ }\textbf {\bibinfo
  {volume} {937}} (\bibinfo {year} {2018})}\BibitemShut {NoStop}%
\bibitem [{\citenamefont {Arrington}\ \emph {et~al.}(2022)\citenamefont
  {Arrington} \emph {et~al.}}]{Harnik}%
  \BibitemOpen
  \bibfield  {author} {\bibinfo {author} {\bibfnamefont {J.}~\bibnamefont
  {Arrington}} \emph {et~al.} (\bibinfo {collaboration} {Fermilab Booster
  Replacement Science Working Group}),\ }\bibfield  {title} {\bibinfo {title}
  {{Physics Opportunities for the Fermilab Booster Replacement}},\ }\href@noop
  {} {\bibfield  {journal} {\bibinfo  {journal} {FERMILAB-FN-1145}\ } (\bibinfo
  {year} {2022})}\BibitemShut {NoStop}%
\bibitem [{\citenamefont {Abusalma}\ \emph {et~al.}(2018)\citenamefont
  {Abusalma} \emph {et~al.}}]{mu2e2}%
  \BibitemOpen
  \bibfield  {author} {\bibinfo {author} {\bibfnamefont {F.}~\bibnamefont
  {Abusalma}} \emph {et~al.},\ }\bibfield  {title} {\bibinfo {title}
  {{Expression of Interest for Evolution of the Mu2e Experiment}},\ }\href
  {https://arxiv.org/abs/1802.02599} {\bibfield  {journal} {\bibinfo  {journal}
  {FERMILAB-FN-1052}\ } (\bibinfo {year} {2018})}\BibitemShut {NoStop}%
\bibitem [{\citenamefont {Acciarri}\ \emph {et~al.}(2015)\citenamefont
  {Acciarri} \emph {et~al.}}]{SBND}%
  \BibitemOpen
  \bibfield  {author} {\bibinfo {author} {\bibfnamefont {R.}~\bibnamefont
  {Acciarri}} \emph {et~al.},\ }\bibfield  {title} {\bibinfo {title} {{A
  Proposal for a Three Detector Short-Baseline Neutrino Oscillation Program in
  the Fermilab Booster Neutrino Beam}},\ }\href
  {https://arxiv.org/abs/1503.01520} {\bibfield  {journal} {\bibinfo  {journal}
  {arXiv:1503.01520}\ } (\bibinfo {year} {2015})}\BibitemShut {NoStop}%
\bibitem [{\citenamefont {Eldred}\ \emph {et~al.}(2021)\citenamefont {Eldred}
  \emph {et~al.}}]{ShiltsevEldred}%
  \BibitemOpen
  \bibfield  {author} {\bibinfo {author} {\bibfnamefont {J.}~\bibnamefont
  {Eldred}} \emph {et~al.},\ }\bibfield  {title} {\bibinfo {title} {{Beam
  intensity effects in Fermilab Booster synchrotron}},\ }\href
  {https://journals.aps.org/prab/abstract/10.1103/PhysRevAccelBeams.24.044001}
  {\bibfield  {journal} {\bibinfo  {journal} {Phys. Rev. Accel. Beams}\
  }\textbf {\bibinfo {volume} {24}} (\bibinfo {year} {2021})}\BibitemShut
  {NoStop}%
\bibitem [{\citenamefont {Foster}\ \emph {et~al.}(2002)\citenamefont {Foster}
  \emph {et~al.}}]{PDriver}%
  \BibitemOpen
  \bibfield  {author} {\bibinfo {author} {\bibfnamefont {G.~W.}\ \bibnamefont
  {Foster}} \emph {et~al.},\ }\bibfield  {title} {\bibinfo {title} {{P}roton
  {D}river {S}tudy. {II}. ({P}art 1)},\ }\href
  {http://lss.fnal.gov/archive/test-tm/2000/fermilab-tm-2169.pdf} {\bibfield
  {journal} {\bibinfo  {journal} {FERMILAB-TM-2169C}\ } (\bibinfo {year}
  {2002})}\BibitemShut {NoStop}%
\bibitem [{\citenamefont {Holmes}\ \emph {et~al.}(2010)\citenamefont {Holmes}
  \emph {et~al.}}]{ICD2}%
  \BibitemOpen
  \bibfield  {author} {\bibinfo {author} {\bibfnamefont {S.~D.}\ \bibnamefont
  {Holmes}} \emph {et~al.},\ }\bibfield  {title} {\bibinfo {title} {{P}roject
  {X} {I}nitial {C}onfiguration {D}ocument - 2},\ }\href
  {http://projectx-docdb.fnal.gov/cgi-bin/RetrieveFile?docid=230} {\bibfield
  {journal} {\bibinfo  {journal} {Project X-DOC-230-v10}\ } (\bibinfo {year}
  {2010})}\BibitemShut {NoStop}%
\bibitem [{\citenamefont {Nagaitsev}\ and\ \citenamefont
  {Lebedev}(2019)}]{Nagaitsev}%
  \BibitemOpen
  \bibfield  {author} {\bibinfo {author} {\bibfnamefont {S.}~\bibnamefont
  {Nagaitsev}}\ and\ \bibinfo {author} {\bibfnamefont {V.}~\bibnamefont
  {Lebedev}},\ }\bibfield  {title} {\bibinfo {title} {{A} {C}ost-{E}ffective
  {R}apid-{C}ycling {S}ynchrotron},\ }\href
  {https://doi.org/10.1142/S1793626819300135} {\bibfield  {journal} {\bibinfo
  {journal} {Reviews of Accelerator Science and Technology}\ }\textbf {\bibinfo
  {volume} {10}},\ \bibinfo {pages} {245} (\bibinfo {year} {2019})}\BibitemShut
  {NoStop}%
\bibitem [{\citenamefont {Eldred}()}]{Eldred2}%
  \BibitemOpen
  \bibfield  {author} {\bibinfo {author} {\bibfnamefont {J.}~\bibnamefont
  {Eldred}},\ }\bibfield  {title} {\bibinfo {title} {{N}ovel {A}pproaches to
  {H}igh-{P}ower {P}roton {B}eams},\ }\href@noop {} {\bibinfo  {journal} {1ST
  International Workshop on Neutrinos from Accelerators (NuFACT2019)}\
  }\BibitemShut {NoStop}%
\bibitem [{LBN(2015)}]{LBNFcdr3}%
  \BibitemOpen
\bibfield  {journal} {  }\bibfield  {title} {\bibinfo {title} {{LBNF/DUNE CDR
  Volume 3: The Long-Baseline Neutrino Facility for DUNE}},\ }\href
  {https://docs.dunescience.org/cgi-bin/ShowDocument?docid=182} {\bibfield
  {journal} {\bibinfo  {journal} {DUNE-doc-182}\ } (\bibinfo {year}
  {2015})}\BibitemShut {NoStop}%
\bibitem [{Ann(2015)}]{Annex3A}%
  \BibitemOpen
  \bibfield  {title} {\bibinfo {title} {{Long-Baseline Neutrino Facility
  (LBNF)/DUNE Conceptual Design Report Annex 3A: Beamline at the Near Site}},\
  }\href {https://lbne2-docdb.fnal.gov/cgi-bin/ShowDocument?docid=10686}
  {\bibfield  {journal} {\bibinfo  {journal} {LBNE-doc-10686-v10}\ } (\bibinfo
  {year} {2015})}\BibitemShut {NoStop}%
\bibitem [{\citenamefont {Fields}(2018)}]{Fields}%
  \BibitemOpen
  \bibfield  {author} {\bibinfo {author} {\bibfnamefont {L.}~\bibnamefont
  {Fields}} (\bibinfo {collaboration} {DUNE collaboration}),\ }\bibfield
  {title} {\bibinfo {title} {{Optimization of the LBNF Neutrino Beam}},\ }\href
  {https://indico.fnal.gov/event/15204/contributions/30218/attachments/18904/23699/Fields-LBNFDUNEBeamOptimization-June2018.pdf}
  {\bibfield  {journal} {\bibinfo  {journal} {High Power Targetry Workshop}\ }
  (\bibinfo {year} {2018})}\BibitemShut {NoStop}%
\bibitem [{Ann(2017)}]{Annex3Aopt}%
  \BibitemOpen
  \bibfield  {title} {\bibinfo {title} {{Long-Baseline Neutrino Facility
  (LBNF)/DUNE Conceptual Design Report Annex 3A Opt: Optimized Neutrino
  Beamline}},\ }\href
  {https://docs.dunescience.org/cgi-bin/sso/ShowDocument?docid=4559} {\bibfield
   {journal} {\bibinfo  {journal} {DUNE-doc-4559-v12}\ } (\bibinfo {year}
  {2017})}\BibitemShut {NoStop}%
\bibitem [{\citenamefont {Densham}(2019)}]{Densham}%
  \BibitemOpen
  \bibfield  {author} {\bibinfo {author} {\bibfnamefont {C.}~\bibnamefont
  {Densham}} (\bibinfo {collaboration} {LBNF collaboration}),\ }\bibfield
  {title} {\bibinfo {title} {{Design Studies of the LBNF/DUNE Target}},\ }\href
  {https://indico.cern.ch/event/773605/contributions/3518761/} {\bibfield
  {journal} {\bibinfo  {journal} {NuFACT'19}\ } (\bibinfo {year}
  {2019})}\BibitemShut {NoStop}%
\bibitem [{\citenamefont {Senor}(2019)}]{Senor}%
  \BibitemOpen
  \bibfield  {author} {\bibinfo {author} {\bibfnamefont {D.}~\bibnamefont
  {Senor}} (\bibinfo {collaboration} {RaDIATE collaboration}),\ }\bibfield
  {title} {\bibinfo {title} {{Radiation Damage Experiments Update from the
  RaDIATE Collaboration}},\ }\href
  {https://indico.cern.ch/event/773605/contributions/3518773/} {\bibfield
  {journal} {\bibinfo  {journal} {NuFACT'19}\ } (\bibinfo {year}
  {2019})}\BibitemShut {NoStop}%
\bibitem [{\citenamefont {Ritz}\ \emph {et~al.}(2014)\citenamefont {Ritz} \emph
  {et~al.}}]{2014p5}%
  \BibitemOpen
  \bibfield  {author} {\bibinfo {author} {\bibfnamefont {S.}~\bibnamefont
  {Ritz}} \emph {et~al.} (\bibinfo {collaboration} {HEPAP Subcommittee}),\
  }\href@noop {} {\emph {\bibinfo {title} {{{B}uilding for {D}iscovery:
  {S}trategic {P}lan for {U}.{S}. {P}article {P}hysics in the {G}lobal
  {C}ontext}}}},\ \bibinfo {type} {Tech. Rep.}\ (\bibinfo {year}
  {2014})\BibitemShut {NoStop}%
\bibitem [{pip(2015)}]{pip2mns}%
  \BibitemOpen
  \href@noop {} {\bibinfo {title} {{PIP-II} {M}ission {N}eed {S}tatemet}},\
  \bibinfo {howpublished}
  {\url{https://pip2-docdbcert.fnal.gov/cgi-bin/cert/ShowDocument?docid=152}}
  (\bibinfo {year} {2015})\BibitemShut {NoStop}%
\bibitem [{pip(2020)}]{pip2pdr}%
  \BibitemOpen
  \href@noop {} {\bibinfo {title} {{PIP-II} {P}reliminary {D}esign {R}eport}},\
  \bibinfo {howpublished}
  {\url{https://pip2-docdb.fnal.gov/cgi-bin/private/ShowDocument?docid=2261}}
  (\bibinfo {year} {2020})\BibitemShut {NoStop}%
\bibitem [{\citenamefont {Antipov}\ \emph {et~al.}(2017)\citenamefont {Antipov}
  \emph {et~al.}}]{IOTA1}%
  \BibitemOpen
  \bibfield  {author} {\bibinfo {author} {\bibfnamefont {S.}~\bibnamefont
  {Antipov}} \emph {et~al.} (\bibinfo {collaboration} {FAST/IOTA
  collaboration}),\ }\bibfield  {title} {\bibinfo {title} {{IOTA (Integrable
  Optics Test Accelerator): facility and experimental beam physics program}},\
  }\href {https://iopscience.iop.org/article/10.1088/1748-0221/12/03/T03002}
  {\bibfield  {journal} {\bibinfo  {journal} {Journal of Instrumentation}\
  }\textbf {\bibinfo {volume} {12}}\bibinfo  {number} { (T03002)}}\BibitemShut
  {NoStop}%
\bibitem [{\citenamefont {Wurtele}\ \emph {et~al.}(2021)\citenamefont {Wurtele}
  \emph {et~al.}}]{IOTA2}%
  \BibitemOpen
\bibfield  {number} {  }\bibfield  {author} {\bibinfo {author} {\bibfnamefont
  {J.~S.}\ \bibnamefont {Wurtele}} \emph {et~al.} (\bibinfo {collaboration}
  {FAST/IOTA collaboration}),\ }\bibfield  {title} {\bibinfo {title}
  {{IOTA-FAST A Leading US Facility for Beam Physics and Accelerator Technology
  R\&D: Long-Term Research Opportunities}},\ }\href
  {https://iopscience.iop.org/article/10.1088/1748-0221/12/03/T03002}
  {\bibfield  {journal} {\bibinfo  {journal} {Snowmass'21 Letter of Interest}\
  } (\bibinfo {year} {2021})}\BibitemShut {NoStop}%
\bibitem [{\citenamefont {Eldred}\ and\ \citenamefont
  {Valishev}(2018)}]{EldredI}%
  \BibitemOpen
  \bibfield  {author} {\bibinfo {author} {\bibfnamefont {J.}~\bibnamefont
  {Eldred}}\ and\ \bibinfo {author} {\bibfnamefont {A.}~\bibnamefont
  {Valishev}},\ }\bibfield  {title} {\bibinfo {title} {{Simulation of
  Integrable Synchrotron with Space-charge and Chromatic Tune-shifts}},\ }\href
  {https://inspirehep.net/literature/1690656} {\bibfield  {journal} {\bibinfo
  {journal} {proceedings of IPAC'18}\ } (\bibinfo {year} {2018})}\BibitemShut
  {NoStop}%
\bibitem [{\citenamefont {Carneiro}(2007)}]{Carneiro}%
  \BibitemOpen
  \bibfield  {author} {\bibinfo {author} {\bibfnamefont {J.-P.}\ \bibnamefont
  {Carneiro}},\ }\bibfield  {title} {\bibinfo {title} {{H- Stripping Equations
  and Application to the High Intensity Neutrino Source}},\ }\href
  {https://beamdocs.fnal.gov/AD/DocDB/0027/002740/001/fnal_2740.pdf} {\bibfield
   {journal} {\bibinfo  {journal} {Beams-doc-2740}\ } (\bibinfo {year}
  {2007})}\BibitemShut {NoStop}%
\bibitem [{\citenamefont {Raparia}(2005)}]{Raparia}%
  \BibitemOpen
  \bibfield  {author} {\bibinfo {author} {\bibfnamefont {D.}~\bibnamefont
  {Raparia}} (\bibinfo {collaboration} {SNS collaboration}),\ }\bibfield
  {title} {\bibinfo {title} {{SNS Injection and Extraction Devices}},\ }\href
  {https://accelconf.web.cern.ch/p05/PAPERS/TOAA007.PDF} {\bibfield  {journal}
  {\bibinfo  {journal} {proceedings of PAC'05}\ } (\bibinfo {year}
  {2005})}\BibitemShut {NoStop}%
\bibitem [{\citenamefont {Hotchi}\ \emph {et~al.}(2017)\citenamefont {Hotchi}
  \emph {et~al.}}]{Hotchi}%
  \BibitemOpen
  \bibfield  {author} {\bibinfo {author} {\bibfnamefont {H.}~\bibnamefont
  {Hotchi}} \emph {et~al.},\ }\bibfield  {title} {\bibinfo {title}
  {{Achievement of a low-loss 1-MW beam operation in the 3-GeV rapid cycling
  synchrotron of the Japan Proton Accelerator Research Complex}},\ }\href
  {https://journals.aps.org/prab/pdf/10.1103/PhysRevAccelBeams.20.060402}
  {\bibfield  {journal} {\bibinfo  {journal} {Phys. Rev. Accel. Beams}\
  }\textbf {\bibinfo {volume} {20}} (\bibinfo {year} {2017})}\BibitemShut
  {NoStop}%
\bibitem [{\citenamefont {Cousineau}\ \emph {et~al.}(2017)\citenamefont
  {Cousineau} \emph {et~al.}}]{Cousineau}%
  \BibitemOpen
  \bibfield  {author} {\bibinfo {author} {\bibfnamefont {S.}~\bibnamefont
  {Cousineau}} \emph {et~al.},\ }\bibfield  {title} {\bibinfo {title} {{High
  efficiency laser-assisted H- charge exchange for microsecond duration
  beams}},\ }\href
  {https://journals.aps.org/prab/abstract/10.1103/PhysRevAccelBeams.20.120402}
  {\bibfield  {journal} {\bibinfo  {journal} {Phys. Rev. Accel. Beams}\
  }\textbf {\bibinfo {volume} {20}} (\bibinfo {year} {2017})}\BibitemShut
  {NoStop}%
\bibitem [{\citenamefont {Tan}\ \emph {et~al.}(2020)\citenamefont {Tan},
  \citenamefont {Bhat},\ and\ \citenamefont {Pellico}}]{Tan}%
  \BibitemOpen
  \bibfield  {author} {\bibinfo {author} {\bibfnamefont {C.~Y.}\ \bibnamefont
  {Tan}}, \bibinfo {author} {\bibfnamefont {C.}~\bibnamefont {Bhat}},\ and\
  \bibinfo {author} {\bibfnamefont {W.}~\bibnamefont {Pellico}},\ }\bibfield
  {title} {\bibinfo {title} {{The required number of wide bore cavities for
  PIPII}},\ }\href
  {https://beamdocs.fnal.gov/cgi-bin/sso/ShowDocument?docid=7879} {\bibfield
  {journal} {\bibinfo  {journal} {Beams-doc-7879-v2}\ } (\bibinfo {year}
  {2020})}\BibitemShut {NoStop}%
\bibitem [{\citenamefont {Hassan}(2015)}]{Hassan}%
  \BibitemOpen
  \bibfield  {author} {\bibinfo {author} {\bibfnamefont {M.}~\bibnamefont
  {Hassan}},\ }\bibfield  {title} {\bibinfo {title} {{Electromagnetic Modeling
  of Fermilab's Booster Cavity}},\ }\href
  {http://beamdocs.fnal.gov/AD/DocDB/0050/005013/002/10-0-TKK_MAH_Booster_RF_Cavity_Replacement__Workshop.pdf}
  {\bibfield  {journal} {\bibinfo  {journal} {FERMILAB-Beams-doc-5013-v2}\ }
  (\bibinfo {year} {2015})}\BibitemShut {NoStop}%
\bibitem [{CBE(2016)}]{CBETA}%
  \BibitemOpen
  \bibfield  {title} {\bibinfo {title} {{Halbach Magnets for CBETA}},\ }\href
  {https://www.classe.cornell.edu/CBETA_PM/161220_review_buildability/reading_material/Halbach_Magnets_Review_Dec17_2016.pdf}
  {\bibfield  {journal} {\bibinfo  {journal} {CBETA Magnet Buildability
  Review}\ } (\bibinfo {year} {2016})}\BibitemShut {NoStop}%
\bibitem [{\citenamefont {Blackmore}(1985)}]{Blackmore}%
  \BibitemOpen
  \bibfield  {author} {\bibinfo {author} {\bibfnamefont {E.~W.}\ \bibnamefont
  {Blackmore}},\ }\bibfield  {title} {\bibinfo {title} {{Radiation Effects of
  Protons on Samarium-Cobalt Permanent Magnet}},\ }\href
  {https://accelconf.web.cern.ch/p85/PDF/PAC1985_3669.pdf} {\bibfield
  {journal} {\bibinfo  {journal} {IEEE TNS}\ }\textbf {\bibinfo {volume}
  {NS-32}} (\bibinfo {year} {1985})}\BibitemShut {NoStop}%
\bibitem [{\citenamefont {Kinsho}\ \emph {et~al.}(2005)\citenamefont {Kinsho},
  \citenamefont {Ogiwara},\ and\ \citenamefont {Saito}}]{Kinsho}%
  \BibitemOpen
  \bibfield  {author} {\bibinfo {author} {\bibfnamefont {M.}~\bibnamefont
  {Kinsho}}, \bibinfo {author} {\bibfnamefont {N.}~\bibnamefont {Ogiwara}},\
  and\ \bibinfo {author} {\bibfnamefont {Y.}~\bibnamefont {Saito}},\ }\bibfield
   {title} {\bibinfo {title} {{Alumina Ceramics Vacuum Duct for the 3GeV-RCS of
  the J-PARC}},\ }\href {https://accelconf.web.cern.ch/p05/PAPERS/RPPE039.PDF}
  {\bibfield  {journal} {\bibinfo  {journal} {proceedings of PAC'05}\ }
  (\bibinfo {year} {2005})}\BibitemShut {NoStop}%
\end{thebibliography}%


  



\end{document}